\documentclass[prd,twocolumn,amsmath,amssymb,floatfix,superscriptaddress,nofootinbib]{revtex4-1}

\usepackage{graphicx}
\usepackage{scrextend}
\usepackage{amssymb}
\usepackage{amsmath}
\usepackage{bm}
\usepackage{color}
\usepackage{multirow}

\DeclareMathAlphabet\mathbfcal{OMS}{cmsy}{b}{n}
\definecolor{darkgreen}{RGB}{50,150,0}
\definecolor{purple}{cmyk}{0.5,1.0,0,0}

\def\edth{\;\raise1.0pt\hbox{$'$}\hskip-6pt\partial}
\def\baredth{\;\overline{\raise1.0pt\hbox{$'$}\hskip-6pt
\partial}}

\newcommand{\Cov}{\mathrm{Cov}}
\newcommand{\PP}{{\phi\phi}}
\newcommand{\ourTh}{\Psi}
\def\be{\begin{equation}}
\def\ee{\end{equation}}
\def\ben{\begin{equation} \nonumber}
\def\een{\end{equation}}
\def\ban{\begin{eqnarray*}}
\def\ean{\end{eqnarray*}}
\def\ba{\begin{eqnarray}}
\def\ea{\end{eqnarray}}
\def\({\left(}
\def\){\right)}

\newcommand{\LCDM}{$\Lambda$CDM }

\newcommand{\Comment}[1]{{}}

\definecolor{ultramarine}{rgb}{0.07, 0.04, 0.56}
\definecolor{cadmiumgreen}{rgb}{0.0, 0.42, 0.24}
\definecolor{indigo(dye)}{rgb}{0.0, 0.25, 0.42}
\usepackage[linktocpage=true]{hyperref}
\hypersetup{
colorlinks=true,
citecolor=ultramarine,
linkcolor=cadmiumgreen,
urlcolor=indigo(dye),
pdfauthor={},
pdftitle={},
pdfsubject={}
}

\begin{document}

\title{CMB Lens Sample Covariance and Consistency Relations}

\author{Pavel Motloch}
\affiliation{Kavli Institute for Cosmological Physics, Department of Physics, University of Chicago, Chicago, Illinois 60637, U.S.A}

\author{Wayne Hu}
\affiliation{Kavli Institute for Cosmological Physics, Department of Astronomy \& Astrophysics,  Enrico Fermi Institute, University of Chicago, Chicago, Illinois 60637, U.S.A}

\author{Aur{\'e}lien Benoit-L{\'e}vy}
\affiliation{Sorbonne Universit{\'e}s, UPMC Univ Paris 06, UMR 7095, Institut d'Astrophysique de Paris, F-75014, Paris, France}

\begin{abstract}
\noindent
Gravitational lensing information from the two and higher point statistics of the CMB temperature and
polarization fields are intrinsically correlated because {they} are lensed by the same
{realization of structure between last scattering  and observation.}
  Using an analytic model for lens sample covariance, we show that there is one mode, separately measurable in
the lensed CMB power spectra and lensing reconstruction, that carries most of this correlation.   Once these measurements become lens sample variance dominated, 
this mode should provide 
a useful consistency check between the observables
that is largely free of sampling and cosmological parameter errors.  Violations of 
consistency could indicate systematic errors in the data and lens reconstruction or
new physics at last scattering, any of which could bias  cosmological inferences and 
delensing for gravitational waves.  
A second mode provides a weaker consistency check for
a spatially flat universe.  Our  analysis isolates the additional information supplied by  lensing in a model independent manner but is also useful for understanding and forecasting CMB cosmological
parameter errors in the extended \LCDM parameter space of dark energy, curvature
and massive neutrinos.  We introduce and test a simple but accurate forecasting technique for this purpose that neither double counts lensing information nor neglects lensing in the observables. 
\end{abstract}

\maketitle

\section{Introduction}
\label{sec:intro}

Power spectra of the cosmic microwave background (CMB) anisotropies have been extremely
valuable in helping to confirm predictions of the standard $\Lambda$CDM cosmological model
and constrain values of cosmological parameters \cite{Ade:2015xua}. Only recently has gravitational lensing of the CMB been detected, first through cross-correlation with galaxy surveys \cite{Smith:2007rg, Hirata08,Hanson:2013hsb}, and then by internal correlations of
the temperature ($T$) \cite{Das:2011ak, Keisler:2011aw, Planck2013XVII}, and polarization
($E$,$B$) \cite{Keisler:2015hfa,Ade:2015zua, Array:2016afx, Sherwin2016} fields, adding a new source of
cosmological information. This secondary signal depends on growth of structure in the
universe, which can be leveraged to break certain parameter degeneracies
in the CMB data and used to better constrain sum of neutrino masses and other parameters
in models beyond $\Lambda$CDM (see \cite{Lewis:2006fu} for a review). 

Information carried by the lensing potential $\phi$ can be recovered either by measuring its effect
on CMB power spectra, in particular the smoothing of the acoustic peaks \cite{Seljak:1995ve} or by measuring
four point functions of the temperature and polarization maps.  The latter is possible, because
gravitational lensing generates a correlation between measured CMB fields and their
gradients \cite{Zaldarriaga:2000ud,Hu:2001fa,Hu:2001kj}, modifying the simple Gaussian statistics of the unlensed CMB.
This non-Gaussian structure can be used to measure the lensing potential, for
example using a quadratic reconstruction \cite{Okamoto:2003zw} or iterative delensing
\cite{Hirata:2003ka,Smith:2010gu}. The reconstructed potential then serves as a new
cosmological observable.

The same non-Gaussianity that makes lensing reconstruction possible is responsible for
correlating the CMB observables and complicates their
analysis.  Gravitational lensing induces nontrivial covariances
between the lensed temperature and polarization data \cite{BenoitLevy:2012va,
Schmittfull:2013uea}. Neglecting these covariances can affect parameter forecasts of
future experiments and analysis of their data.   

In particular, future experiments are expected to have their lensing information limited by sample
variance of the lenses: the fact that on the same patch of sky, the gravitational lensing of
all CMB observables is due to the same realizations of a finite number of lens modes.
 In this work we use an extension of the 
analytical model of Ref.~\cite{BenoitLevy:2012va}
  to include covariances between power spectra $C^{XY}_\ell$ of the lensed
CMB temperature and polarizations with the power spectra of the reconstructed lensing
potential, recently also discussed in \cite{Green:2016cjr, Peloton:2016kbw}. 
With this model we then investigate how these covariances affect parameter forecasts
and construct sharp consistency relations between the two types of observables that can be used
to test for foregrounds, systematics or new physics.

The outline of the paper is as follows.  
In  \S \ref{sec:model} we present the analytical model for lens sample covariances. 
We analyze their impact on cosmological parameters in \S \ref{sec:parameters} and separate 
information on them into lensing and non-lensing based sources.  
Based on this separation, we determine the modes that most strongly covary between CMB power
spectra and lens reconstruction in \S \ref{sec:kl}.  These provide consistency relations between observables that
are largely immune to lens sample variance and cosmological parameter uncertainties.
We discuss these results in \S \ref{sec:discuss}.  In the Appendix we use these results
to develop a new accurate but simple Fisher forecasting technique in the extended \LCDM parameter space
that avoids double counting 
lensing information, and compare
it with other similar but less accurate approaches.

\section{Analytic lens covariance model}
\label{sec:model}

In this section, we present an analytical model describing 
non-Gaussian covariances between the $C_\ell^{xy}$ power and cross spectra observables
induced by gravitational lensing through the same lenses on the sky.
Here these $xy$  spectra are the CMB temperature power spectra $TT$,  $E$-mode
polarization power $EE$,  temperature-polarization cross spectra $TE$,  $B$-mode
polarization power $BB$ and the power spectrum of the  lens potential
$\phi\phi$.

As a notational short  hand, we denote the subset of $xy$ that includes only the
CMB power spectra  with capital letters $XY$: $xy \in \{XY,\phi\phi\}$, whereas $XY\in \{TT,EE,TE,BB\}$.
 Note that although the $T\phi$ and $E\phi$ spectra are also observable we
omit them as a source of information but include them in the covariance of other spectra. We comment more on this choice in \S \ref{sec:parameters}.
Covariances predicted by
this model have been tested against numerical simulations in \cite{BenoitLevy:2012va} for
the $XY$ power spectra; here we use the physical intuition gained in
\cite{BenoitLevy:2012va} to extend the same model to include their covariance with
measurements of $\PP$. A similar model has recently
been also used in \cite{Green:2016cjr, Peloton:2016kbw}. 

\begin{figure}
\center
\includegraphics[width = 0.49 \textwidth]{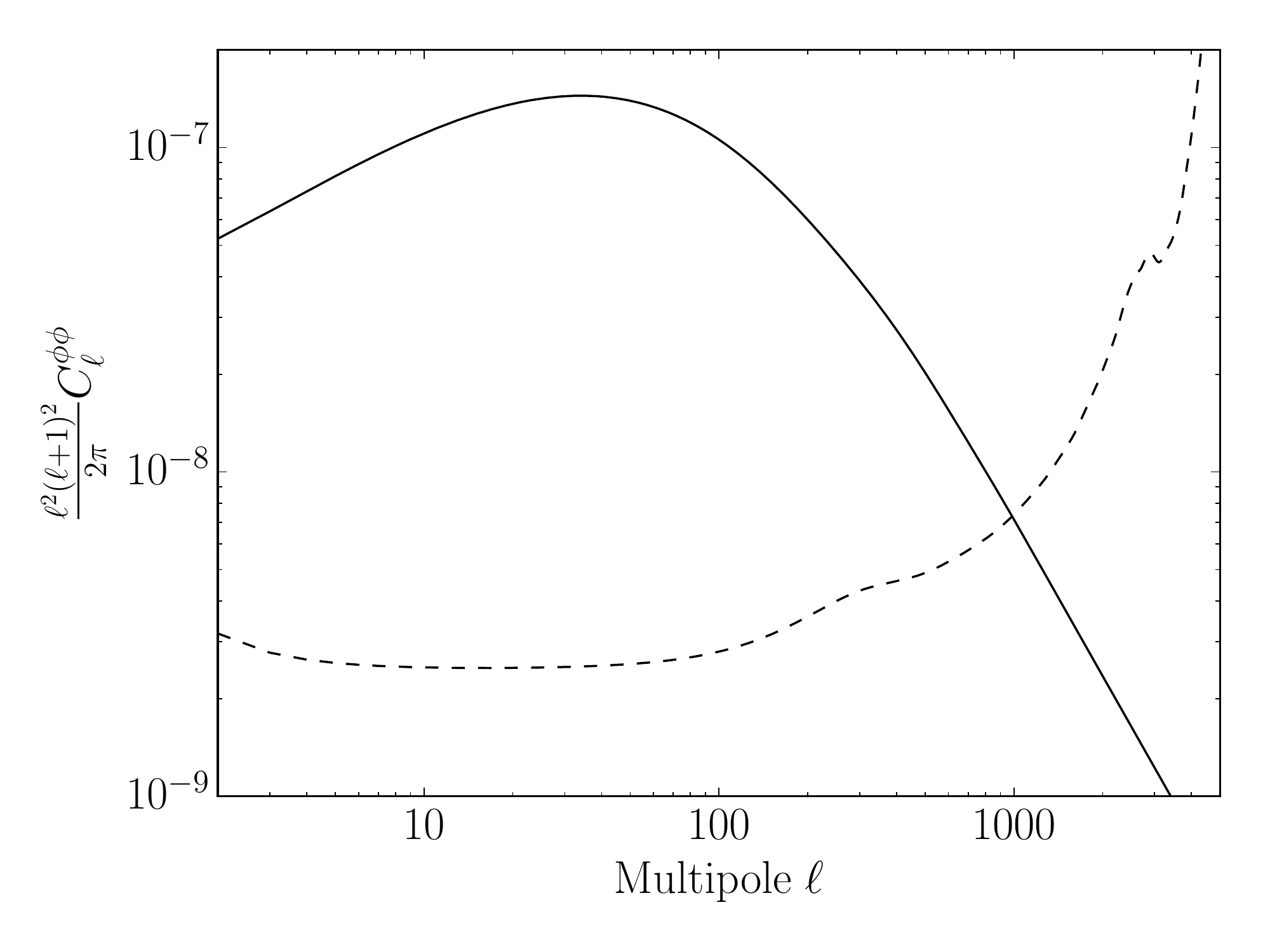}
\caption{Comparison of the lensing potential power spectra $C^\PP_\ell$ (solid) with the
reconstruction noise forecast in this work (dashed, see text for details).  The forecast is lens sample variance
limited for $\ell \lesssim 10^3$.  } 
\label{fig:Npp}
\end{figure}

In this model the correlation matrix is split into  ``Gaussian part'' $\mathcal{G}$ that is diagonal
in multipole space and  $\mathcal{N}$ which describes non-Gaussian correlations between multipoles,
\be
\label{full_covariance}
	\Cov^{xy,wz}_{\ell \ell'} = \mathcal{G}^{xy,wz}_{\ell \ell'}
	+\mathcal{N}^{xy,wz}_{\ell \ell'} .
\ee

The Gaussian part is modelled after the covariance of Gaussian random fields as
\be
\label{gaussian_covariance}
	\mathcal{G}^{xy, wz}_{\ell \ell'} = \frac{\delta_{\ell \ell'}}{2\ell + 1}
	\left[C_{\mathrm{exp},\ell}^{xw}C_{\mathrm{exp},\ell}^{yz} +
		C_{\mathrm{exp},\ell}^{xz} C_{\mathrm{exp},\ell}^{yw}\right] ,
\ee
where the expectation value of the experimentally measured lensed CMB power spectra
$C^{xy}_\mathrm{exp}$ includes the noise power spectrum $N_\ell^{xy}$
\be
	C_{\mathrm{exp},\ell}^{xy} = C_\ell^{xy} + N_\ell^{xy}.
\ee
For noise in temperature and polarizations, we assume a Gaussian noise spectra  \cite{Knox1995}
\be
\label{Gauss_noise}
	N_\ell^{XY} = \Delta_{XY}^2 e^{\ell(\ell+1){\theta_\mathrm{FWHM}^2}/{8 \log 2} },
\ee
where $\Delta_{XY}$ is the instrumental noise (in $\mu$K-radian) and $\theta_\mathrm{FWHM}$ is the beam size
(in radians). 

In this work we investigate a simplified experimental setup of a full sky experiment with 
specifications inspired by  CMB Stage 4 \cite{Abazajian:2016yjj}. We consider a $1'$ beam,
$\Delta_{TT}=1\,\mu$K$'$, $\Delta_{EE}=\Delta_{BB}=1.4\,\mu$K$'$, and $\Delta_{TE}=\Delta_{TB}=\Delta_{EB}=0$ and 
use measurements from $\ell = 2 - 3000$. 
CMB Stage 4 measurements at $\ell>3000$ have negligible impact on our
results (see \S \ref{sec:parameters} and \ref{sec:kl}).

We also assume measurements of $C_\ell^\PP$ from $\ell = 2 - 5000$ with the reconstruction noise $N_\ell^\PP$  of  the minimal variance quadratic estimator \cite{Okamoto:2003zw}, commonly known as $N^{(0)}$ noise bias,  and ignore other noise biases and trispectrum terms \cite{Schmittfull:2013uea} (see \S \ref{sec:discuss}).
Comparison of the $C_\ell^\PP$ with the reconstruction noise for our experiment
$N^\PP_\ell$ is plotted in Figure~\ref{fig:Npp}.  Notice that for these specifications,
the lens reconstruction is sample variance dominated for $\ell \lesssim 10^{3}$. This is
the fundamental assumption underlying this work: that lens sample variance will in the
future dominate the measurements of the lens power spectrum at low multipoles.  The
consistency check proposed in \S \ref{sec:kl} can be viewed as an operational test of this
assumption and we comment more on current simulation-based tests in \S \ref{sec:discuss}.

\begin{figure*}
\center
\includegraphics[width = 0.75 \textwidth]{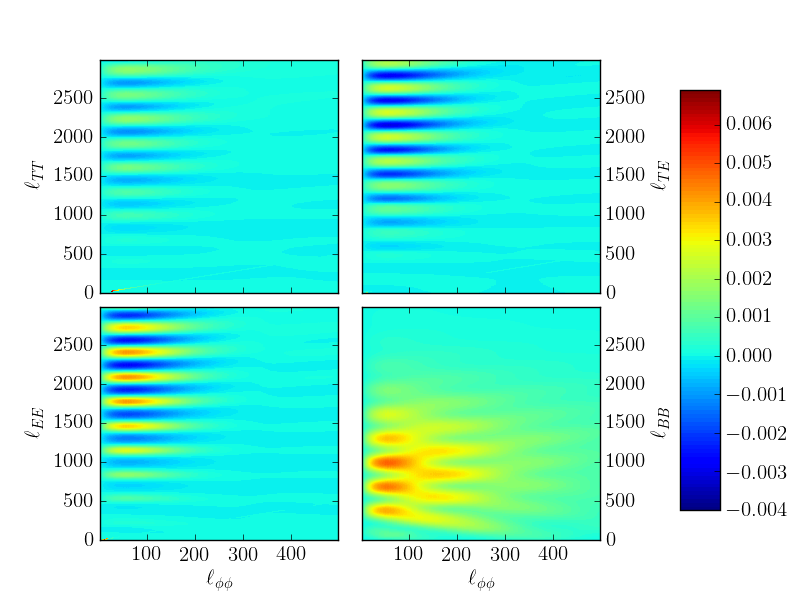}
\caption{Correlation matrix $R^{XY,\PP}_{\ell_{XY},\ell_{\PP}}$ \eqref{scaled_covariance}
between the $C_\ell^{XY}$ CMB power spectra 
and the power spectra of the reconstructed lensing
potential $C_\ell^{\PP}$. Barely visible  features  for $\ell_{XY}=\ell_{\PP} \lesssim 50$ in the first three panels represent contributions
from the Gaussian terms due to nonzero $C_\ell^{T\phi}, C_\ell^{E\phi}$. }
\label{fig:covariances}
\end{figure*}

Even if we assume that the unlensed CMB fields $\tilde X$ and $\phi$ are Gaussian, the
lensed CMB fields $X$ are not. In our model, we take two non-Gaussian terms to compose the
full covariance,
\be
\label{nongauss_covariance}
	\mathcal{N}^{xy,wz}_{\ell\ell'} =  \mathcal{N}^{(\phi)xy,wz}_{\ell\ell'} +
	 \mathcal{N}^{(E)xy,wz}_{\ell\ell'} ,
\ee
which we now describe.

Gravitational lensing induces non-Gaussian covariances between the data because all
power spectra are affected by the same realization of the lensing potential; sample
variance fluctuations of the lensing power produce coherent changes in all the observed
power spectra. The effect accumulates  over the whole multipole range  of the lenses and is largest for
those $C^{XY}_\ell$ which are most strongly affected by lensing. It is modeled by adding an extra term 
\be
\label{nongaussian_covariance}
	 \mathcal{N}^{(\phi)xy,wz}_{\ell \ell'} = \sum_{L}
	\frac{\partial{C_\ell^{xy}}}{\partial C_{L}^{\PP}}\Cov^{\PP}_{LL}
	\frac{\partial{C_{\ell'}^{wz}}}{\partial C_{L}^{\PP}}
\ee
to the non-Gaussian covariance $\mathcal{N}$.
The power spectra derivatives are in practice calculated using a two point central difference scheme
from results obtained using CAMB \cite{Lewis2002}. 
For the reconstructed potential we take
$\mathcal{N}^{(\phi)\PP,\PP}_{\ell\ell'} = 0$ as the corresponding variance is part of the
Gaussian term.

Sample variance of the unlensed $\tilde E \tilde E$ power spectrum and its coherent
propagation into the lensed power spectra through gravitational lensing produces similar but 
typically weaker effects. Following \cite{BenoitLevy:2012va}
we include this contribution only for
$\Cov^{XY,BB}_{\ell\ell'}$ with 
\be
		 \mathcal{N}^{(E)XY,BB}_{\ell, \ell'} = \sum_{L}
		\frac{\partial{C_\ell^{XY}}}{\partial C_{L}^{\tilde X \tilde Y}}\Cov^{\tilde X
		\tilde Y, \tilde E \tilde E}_{L,L}
		\frac{\partial{C_{\ell'}^{BB}}}{\partial C_{L}^{\tilde E \tilde E}}.
\ee
Other sample covariance effects from unlensed fields on $XY$ 
are negligible in comparison \cite{BenoitLevy:2012va}.    We also assume that the analogous
terms involving the reconstruction noise, e.g. $\partial N_l^{\phi\phi}/\partial C_L^{\tilde E\tilde E}$ and other non-Gaussian reconstruction terms are negligible.
This should be a good approximation in the lens sample dominated regime $\ell \lesssim 10^3$ 
(see \S \ref{sec:discuss}).

The covariances $\Cov^{XY,WZ}$ we obtain for the CMB power spectra
qualitatively agree with those plotted in Fig. 1 of
\cite{BenoitLevy:2012va} for the same analytical model for covariances but for a slightly
different cosmological model. The less well studied covariances $\Cov^{XY,\PP}$ 
are shown in Figure~\ref{fig:covariances}; for illustrative purposes we plot the correlation 
coefficient
\be
\label{scaled_covariance}
	R^{XY,\PP}_{\ell \ell'} =
	\frac
	{\Cov^{XY,\PP}_{\ell\ell'}}
	{\sqrt{\Cov^{XY,XY}_{\ell\ell}\Cov^{\PP,\PP}_{\ell'\ell'}}} .
\ee
In this plot we assume experimental and reconstruction noise for our reference experiment.

We see that the covariances peak for $\ell' =\ell_\PP \sim 100-200$ which reflects the fact
that most of the lensing is caused by lenses at these scales. In covariances with $TT, TE$
and $EE$ there are alternating regions of positive and negative correlations,
corresponding to smearing of the peaks and troughs; correlation with $BB$ also shows
acoustic features due to oscillations in the unlensed $C^{\tilde E \tilde E}_\ell$ on top
of a positive definite contribution. The broad band $BB$ power thus coherently covaries
with the lens power \cite{Smith:2004up}. These results also agree with
Ref.~\cite{Green:2016cjr, Peloton:2016kbw}.

\section{Parameter forecasts}
\label{sec:parameters}
In this section we investigate the impact of lens sample
 covariances between measurements of  CMB power spectra  and the  lensing potential  on cosmological parameters.  This impact comes through the additional information that lensing supplies
 on parameters.  We show that to good approximation  this information in the lensed CMB power spectra can be considered independently from that of the
 unlensed CMB power spectra, effectively as direct measurements of the lens power spectrum itself.

\subsection{Cosmological parameters}
\label{sec:standard_fisher}

In this work we focus on  extensions of the standard 6 parameter $\Lambda$CDM
cosmological model which we allow to vary two at a time: the sum of masses of the  neutrino species $\sum m_\nu$,  the dark energy
equation of state $w$, and the spatial curvature $\Omega_K$.  
For the $\Lambda$CDM
parameters we take $\Omega_b h^2$, the physical baryon density;
$\Omega_c h^2$, the physical cold dark matter density; $n_s$, the tilt of the scalar power spectrum; $A_s$, its amplitude; 
and $\tau$, the optical depth to recombination.  We choose $\theta$, the angular scale of the sound horizon
at recombination, as opposed to the
Hubble constant $h$, as the sixth independent parameter 
given the angular diameter distance degeneracy between $h$ and parameters
such as $w$ and $\Omega_K$ in the unlensed CMB.  This choice also improves the numerical
stability of forecasts.
We also assume that tensor modes are negligible so that
there is no unlensed $B$ mode.  We call a set of 8 cosmological parameters of the extended
$\Lambda$CDM family $\theta_A$.  Values of the cosmological  parameters for
the fiducial model used in this work are summarized in Table~\ref{tab:fiducial}.

Our assumptions about measurement noise and characterization of lens sample variance
 in the covariance matrix  are summarized in the previous section. 
In general, we forecast parameter errors
given a covariance matrix of a set of observables $D_i$ using the Fisher matrix
\be
\label{fisher}
	F_{AB} = \sum_{ij} \frac{\partial D_i}{\partial \theta_A} \, \Cov^{-1}_{ij} \,
	\frac{\partial D_j}{\partial \theta_B} .
\ee
The inverse Fisher matrix represents an estimate of the covariance matrix of the parameters
\be
\label{fisher_to_cov}
	\Cov_{\theta_A,\theta_B} = (F_{AB})^{-1} .
\ee
Prior information is included by adding its Fisher matrix before inverting.
\begin{table}
\caption{Fiducial parameters used in the analysis with extensions to the standard
$\Lambda$CDM parameters listed last.} 
\label{tab:fiducial}
\begin{tabular}{c|c}
\hline\hline
Parameter & Fiducial value\\
\hline
$h$ & 0.675\\
$\Omega_c h^2$ & 0.1197 \\
$\Omega_b h^2$ & 0.0222 \\
$n_s$ & 0.9655 \\
$A_s$ & $2.196 \times 10^{-9}$ \\
$\tau$ & 0.06 \\
\hline
$\sum m_\nu$ & 60 meV \\
$w$ & $-1$\\
$\Omega_K$ & 0\\
\hline\hline
\end{tabular}
\end{table}

\begin{figure*}
\center
\includegraphics[width = 0.49 \textwidth]{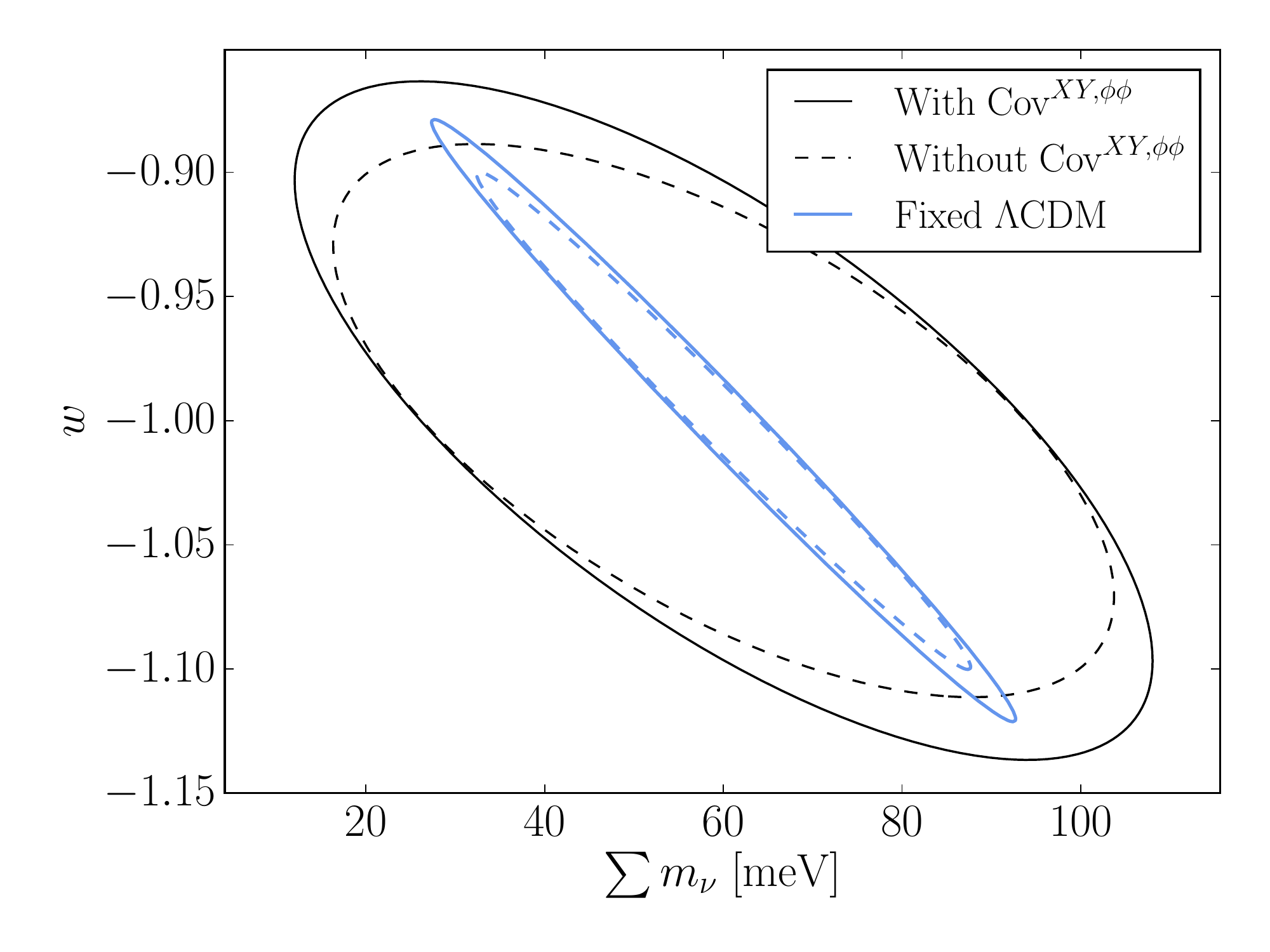}
\includegraphics[width = 0.49 \textwidth]{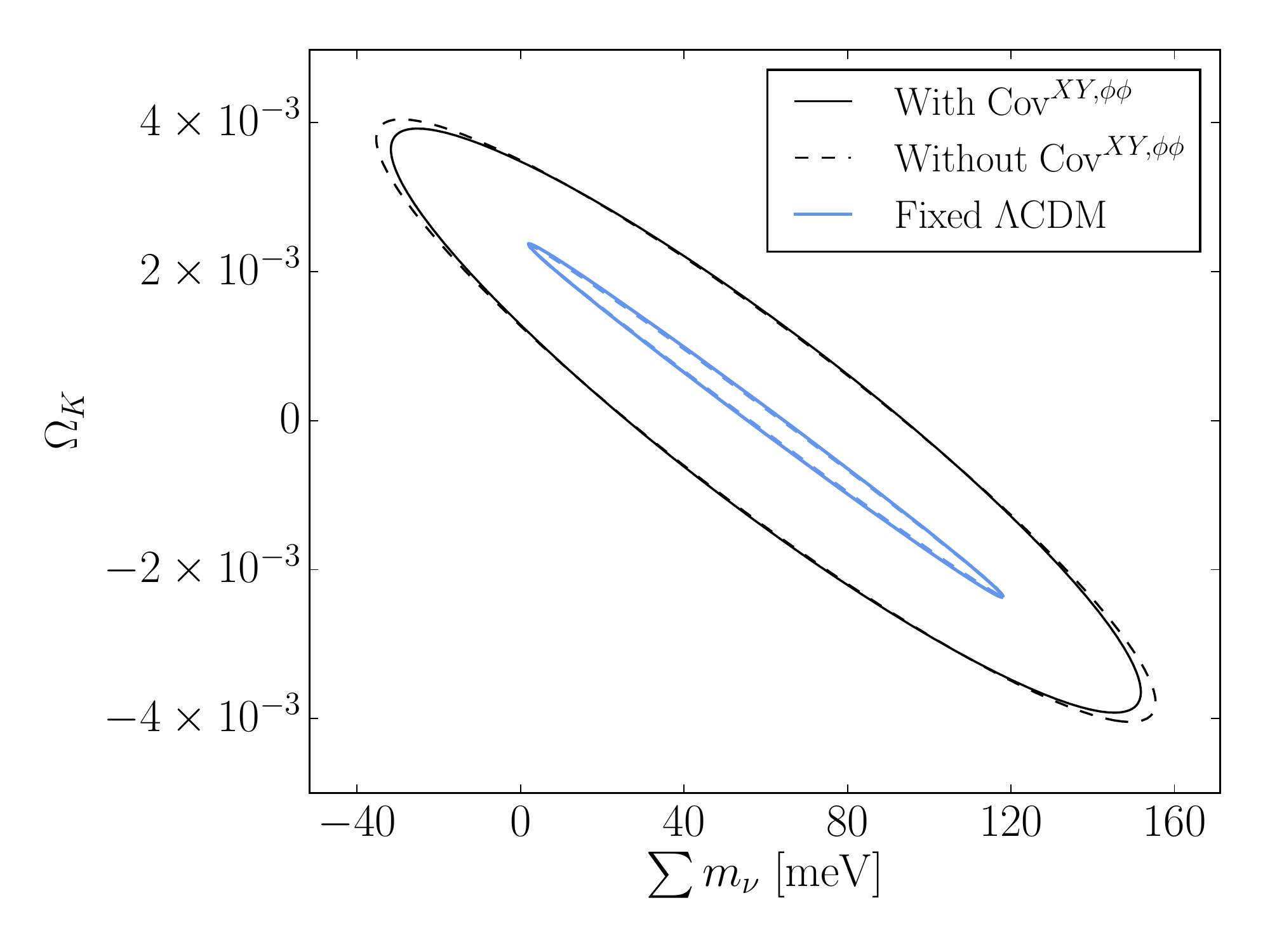}
\caption{Forecasts for 2 parameter extensions to $\Lambda$CDM: $w$-$\sum m_\nu$ (left) and $\Omega_K$-$\sum m_\nu$ (right). Black curves
show $\Delta \chi^2 = 1$ constraints  considering the full covariance (solid)
and with covariances $\Cov^{XY,\PP}$
neglected (dashed);  $\Lambda$CDM parameters are marginalized over. The blue curves
show the same constraints with $\Lambda$CDM parameters fixed to their fiducial values. 
}
\label{fig:param_constraints}
\end{figure*}

In Figure~\ref{fig:param_constraints} we compare how the  Fisher forecasts on the two
extensions of $\Lambda$CDM change  when we
neglect the effect of $\Cov^{XY,\PP}$, the lens sample covariances between the CMB and lens power spectra. In these plots, \LCDM parameters are
marginalized over and the third \LCDM extension fixed. While for
$w$ and $m_\nu$, the effect is sizable and amounts to $\sim 20\%$, for $\Omega_K$ and $m_\nu$ the 
effect is much smaller. These differences reflect parameter degeneracies in the lensing
observables.

We also show in Figure~\ref{fig:param_constraints} the same constraints with the 6
$\Lambda$CDM parameters fixed. It is clear that  the best constrained direction
is limited by parameter degeneracies,  especially with $\Omega_c h^2$
\cite{Smith:2006nk}.  The worst constrained direction is limited instead by the ability of
lensing or other constraints to separate the two additional parameters.

Conversely, in the  $\Lambda$CDM model with only the 6 standard parameters varied, 
parameter errors change by less that 4\% when neglecting
$\Cov^{XY,\PP}$. This reflects the fact that these parameters are well-constrained
even in the absence of lensing.

One of the motivations for the rest of the paper will be to understand these behaviors in
terms of the additional versus redundant information that lensing observables supply. From
the redundant information we will construct sharp consistency tests whose violation would
imply systematic errors or violations of fundamental physical assumptions.

Note also that constraints on cosmological parameters depend strongly on how well $\tau$ is
constrained whereas those on the lensing power spectrum $C_\ell^\PP$ itself
do not \cite{Smith:2006nk}.
For the measurements to cleanly separate $C_\ell^\PP$ information, we
primarily need the unlensed CMB in the acoustic regime $C_\ell^{\tilde X\tilde Y}$ to be
well-characterized. On the other hand, in terms of cosmological parameters, the amplitude of these spectra in this regime is proportional to
$A_s e^{-2\tau}$. The leverage on cosmological parameters gained through comparing
the initial amplitude $A_s$ to the growth-dependent lensing amplitude depends on how well
$\tau$ is measured. In our experimental setup we assumed for simplicity that polarization
information will be obtained for the full range of multipoles $\ell = 2-3000$, which results in a nearly cosmic variance
limited constraint on $\tau$ of  $\sigma(\tau)\approx 0.002$. This is about five times better
than current best constraints from Planck \cite{Adam:2016hgk} and furthermore assumes a
fixed functional form for reionization \cite{Hu:2003gh,Heinrich:2016ojb}. If the final
Planck release does not improve these constraints to substantially below $\sigma(\tau)
\sim 0.01$, this uncertainty will dominate the interpretation of lensing constraints for
cosmological parameters \cite{Smith:2006nk,Allison:2015qca} since it will be difficult to
improve using ground-based instruments. 

More concretely, removing  polarization data from $\ell < 30$ from
our forecasts and replacing it with a prior  of $\sigma_\tau = 0.01$, the
errors in the worst constrained direction in Figure~\ref{fig:param_constraints} do
not significantly change, while those in the best constrained direction degrade
by roughly a factor of two.  On the other hand, characterizing the information on the
power spectrum of the lenses does  not depend strongly on the measurements of $\tau$ and this will be the main
focus of the remainder of this work. 

\subsection{Lens and unlensed information}

CMB information on a given cosmological parameter  comes both from its effect on the
unlensed CMB power spectra $C_\ell^{\tilde X\tilde Y}$ with $\tilde X\tilde Y \in \tilde
T\tilde T,\tilde T\tilde E,\tilde E\tilde E$ and on the
lenses $C_\ell^{\phi\phi}$.  It is conceptually useful to separate these sources of information. 
Indeed, beyond the cosmological parameters considered in the previous section, the total information
in the CMB observables  is carried by all two point
functions for $\tilde T,\tilde E,\tilde B,\phi$, assuming they obey Gaussian statistics; recovery of this complete set of information is
 the ultimate goal of CMB delensing efforts.  By first extracting the lensing information we can also
 further separate the information from lensed CMB power spectra and reconstruction.  
 The latter can be used to form consistency tests between the two sources of lensing information.

Indeed, the Planck satellite found a mild discrepancy between the amount
of lensing present in the $TT$ power spectrum and the $TT$ reconstructed lensing potential
\cite{Aghanim:2015xee}. While these sources of lensing information are still
limited by noise rather than by lens sample variance,
if such discrepancies persist in future experiments, they may indicate
systematic errors in the experiment or the data analysis technique which could obstruct delensing efforts.  
By checking for consistency at the power spectra level, one can provide proof against such problems
before making incorrect cosmological inferences.

In principle the full implementation of this approach would be to consider every multipole in
$C_\ell^{\tilde X \tilde Y}$ and $C_\ell^{\phi\phi}$ as a parameter in its own right.
However, since the high redshift universe is well described by a $\Lambda$CDM-like model, we choose to
parameterize the unlensed power spectra  $C_\ell^{\tilde X \tilde Y}$ in terms of a small number of parameters
$\tilde \theta_A$. These $\tilde \theta_A$ change the unlensed power
spectra in exactly the manner of the $\Lambda$CDM parameters $\theta_A$, but unlike those,
they have no effect on $C_\ell^\PP$. 

The lens power spectrum
is instead described by a more complete set of parameters $p_\alpha$, reflecting the wider
range of possibilities during the acceleration epoch.
For practical reasons, instead of considering each multipole $C_\ell^\PP$ of the lensing
potential as a parameter, we assume that the power spectrum is sufficiently smooth in $\ell$
that we can approximate it with binned perturbations around the fiducial model. We
then define a set of parameters $p_\alpha$ by
\be
	\ln C^\PP_\ell \approx \ln C^\PP_\ell \Big|_\mathrm{fid} + \sum_{\alpha = 1}^{N_\phi} p_\alpha
	B_\alpha^{\phi, \ell} ,
\ee
where $B_\alpha^{\phi, \ell}$ describes the binning and is defined as
\be
	B^{\phi, \ell}_\alpha =
	\begin{cases}
		1 & \ell_\alpha \leq \ell < \ell_{\alpha + 1}\\
		0 & \mathrm{otherwise}
	\end{cases} .
\ee
Expansion in $\ln C_\ell^{\PP}$ is chosen to assure positivity of the power spectrum.
Any cosmological model which predicts a smooth variation of $\ln C^\PP_\ell$ from the fiducial
model can be captured in these parameters as
\be
	p_\alpha = \frac{1}{\Delta \ell_\alpha} \sum_\ell B^{\phi, \ell}_\alpha
	\delta \ln C^\PP_\ell  ,
\ee
where $\Delta \ell_\alpha = \ell_{\alpha + 1} - \ell_\alpha$ is the width of bin $\alpha$. 
We consider uniform binning with bins of width 5 in this paper; we
do not expect binning to have any effect on our conclusions. 
Changes to the lensing potential are allowed up to $\ell = 5000$, given by the $\ell$ range in which we
assume the reconstruction data are measured.

The full set of parameters which we will constrain with a Fisher analysis is then 
\be
	P_\mathrm{tot} = \{\tilde \theta_1, \tilde \theta_2, \dots, p_1, p_2, \dots\} ,
\ee
where $\tilde \theta_A$ only affect the unlensed power spectra and $p_\alpha$ only affect the
lensing potential. A given cosmological parameter $\theta_A$ jointly changes $\tilde
\theta_A$ and $p_\alpha$.

In principle to fully represent a cosmological parameter in this way we would have to
account for the covariance between the lens power spectrum and the unlensed CMB spectra  induced by  $C_\ell^{\tilde T\phi},
C_\ell^{\tilde E\phi}$ -- the ISW-lens and reionization-lens correlations respectively.  We
could in principle add these as parameters to form a complete description. However, these
appear only on the largest, severely cosmic variance limited scales which will also be
difficult to extract due to foregrounds and systematics.  For this reason we completely
neglect them from this section onwards by setting $C_\ell^{T\phi}= C_\ell^{E\phi}=0$
everywhere, which means also in
the Gaussian covariance. We checked that omitting these contributions to the covariance
matrix has only a small effect on parameter constraints in
Figure~\ref{fig:param_constraints}. 

\subsection{Independent approximation}

We can take the lens vs.~unlensed information split of the previous section one step further
and assume that the data constrain parameters of this split independently so that the
$\tilde \theta_A$ and
$p_\alpha$ errors do not covary.  
To the extent that this approximation is true, we can consider the lens information as independent.
Physically, this approximation involves the assumption that changes in the unlensed CMB
and lens
power
spectra do not produce degenerate effects in the lensed CMB.
We can test this approximation by comparing cosmological parameter constraints on $\theta_A$
as constructed from $\tilde \theta_A$ and $p_\alpha$ with the direct forecasts.

Under this approximation we first construct independent Fisher matrices in the $p_\alpha$ space
\ba
\label{F_lenses}
	{F_{\alpha\beta}^\mathrm{lenses} =} 
	&&\sum_{
	\substack{
		\text{$ \ell,\ell'$}
		\\
		\text{$xy, wz$}
	}
	}
	\frac{\partial C^{xy}_\ell}{\partial p_\alpha} 
	\(\Cov^{xy,wz}_{\ell \ell'}\)^{-1}
	\frac{\partial C^{wz}_{\ell'}}{\partial p_\beta}
\ea
with the unlensed CMB spectra $C_\ell^{\tilde X\tilde Y}$ fixed to their fiducial values and the
$\tilde \theta_A$ space 
\ba
	F_{AB}^\mathrm{unl} &=& \sum_{	\substack{
		\text{$ \ell,\ell'$}
		\\
		\text{$XY, WZ$}
	}} 
	\frac{\partial C^{XY}_\ell}{\partial \tilde \theta_A} 
	\(\Cov^{XY,WZ}_{\ell \ell'}\)^{-1}
	\frac{\partial C^{WZ}_{\ell'}}{\partial \tilde \theta_B}
\ea
with $C_\ell^{\phi\phi}$ fixed to their fiducial values.  Note that $\phi\phi$ has no 
dependence on $\tilde\theta_A$ and so those spectra do not enter into the sum.

\begin{figure*}
\center
\includegraphics[width = 0.49 \textwidth]{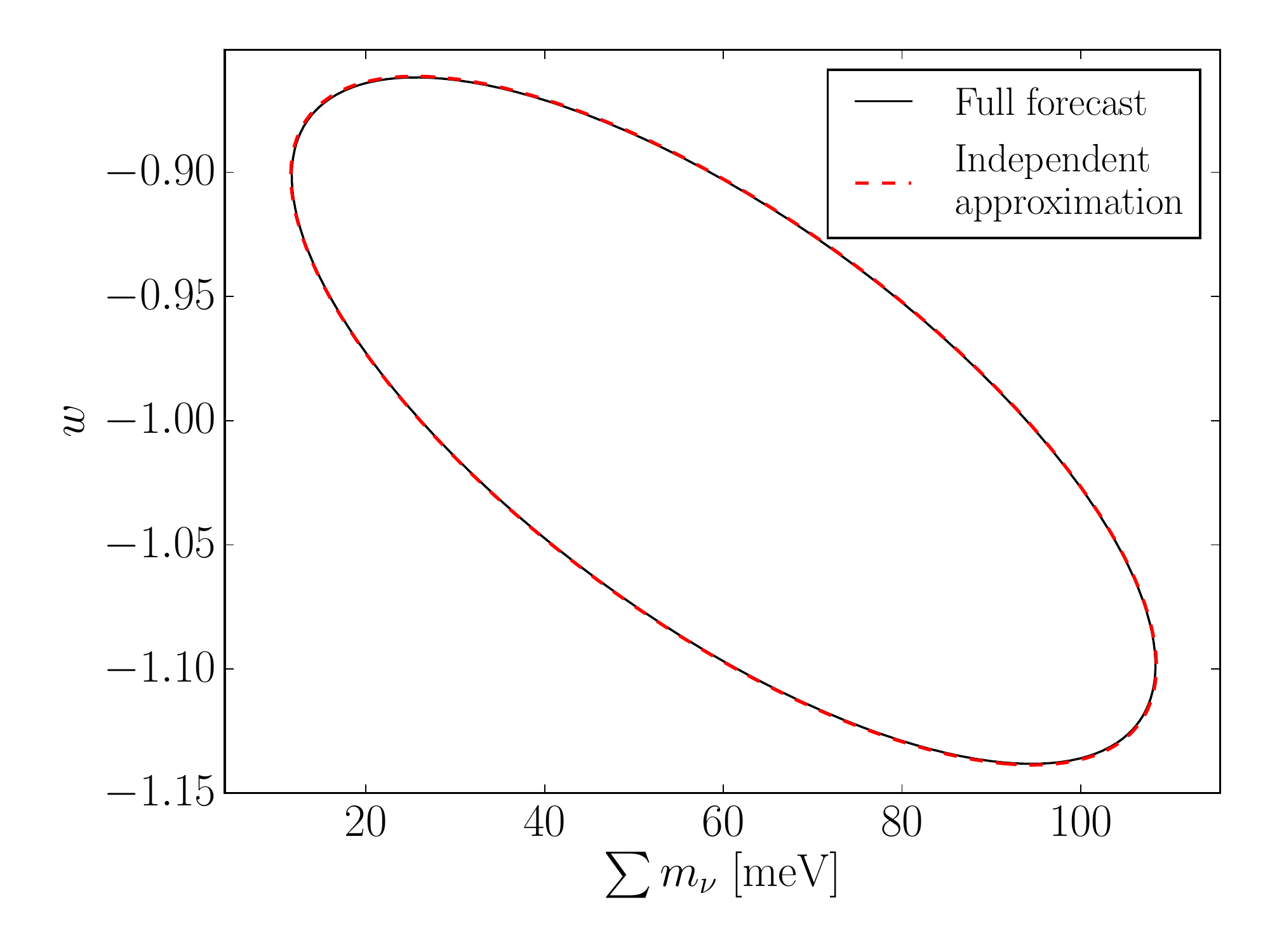}
\includegraphics[width = 0.49 \textwidth]{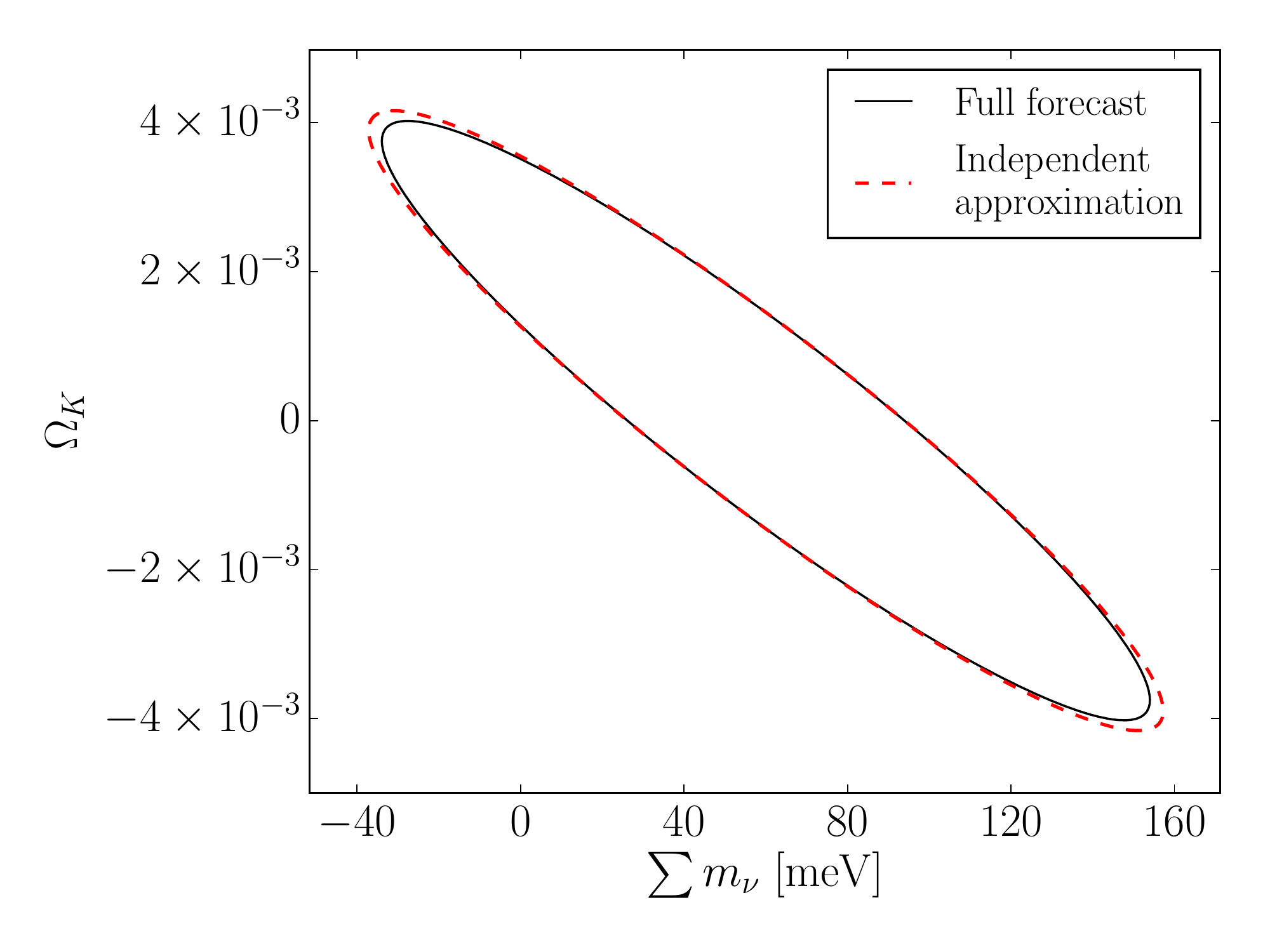}
\caption{Accuracy of the independent lensing information model  of \eqref{simple_model} (red dashed)
compared with the full Fisher forecast for the cases from Fig.~\ref{fig:param_constraints} (black solid). 
}
\label{fig:model_comparison}
\end{figure*}

We can then obtain the total Fisher matrix of the cosmological parameters by the Jacobian
transform
\be
\label{simple_model}
	F_{AB} = F_{AB}^\mathrm{unl} + \sum_{\alpha,\beta} 	
	\frac{\partial p_\alpha}{\partial \theta_A}
 F_{\alpha\beta}^\mathrm{lenses}\frac{\partial p_\beta}{\partial \theta_B}  .
\ee
In Figure~\ref{fig:model_comparison} we compare constraints obtained from the independent model
\eqref{simple_model} with constraints from the full Fisher analysis. We see that the model
indeed works very well and the assumption about independent measurement of the unlensed power
spectra and the lensing potential is justified in these examples.  As we discuss in \S \ref{sec:consistency}, spaces that
involve $\Omega_K$ provide an especially stringent test of the independent approximation.

Because to calculate $F_{AB}^\mathrm{unl}, F_{\alpha\beta}^\mathrm{lenses}$ one needs to know the full
covariance matrix for the lensed observables, this split does not represent any practical simplification
for calculation of the Fisher matrix unlike the related ``additive" approximation in
Ref.~\cite{Smith:2006nk} that utilizes the unlensed spectra as observables.  Conversely,
we do not incur errors from conflating unlensed power spectra with direct observables.
In Appendix~\ref{sec:appendix} we introduce a new forecasting approximation which combines
the virtues of these two approaches: simplicity and accuracy.

\section{Redundancy and Consistency}
\label{sec:kl}

Given the technique for isolating information about the lens power spectrum introduced in the previous section,
we can now assess the level of redundancy and consistency between the information coming
from lensed CMB power spectra and lensing power spectrum.
 This study both helps explain constraints
on cosmological parameters and enables the construction of sharp consistency tests between
these two aspects of lensing in the data that are nearly immune to sample variance.

\subsection{Consistency of covarying modes}
\label{sec:consistency}

We can use the  Karhunen-Lo{\`e}ve (KL) transform\footnote{The KL transform is often used in cosmology to
define signal-to-noise eigenmodes for optimal data compression  \cite{Bond:1994aa,Bunn:1994pi,Vogeley1996}; 
our use follows \cite{Smith:2006nk} in comparing information in two different  covariance matrices.}
 to extract the modes or
linear combinations of the lens parameters $p_\alpha$ that are most impacted by the
$\Cov^{XY,\PP}$ covariance between the measurements of the lensed CMB power
spectra $C_\ell^{XY}$ and the lens power spectra $C_\ell^{\phi\phi}$. These modes carry
redundant information between $XY$ and $\phi\phi$ that can be used as a consistency check
on the data and analysis techniques.

To assess the impact of the $XY,\PP$ covariance, we consider two versions of the inverse Fisher matrix for $p_\alpha$, 
\begin{equation}
\Cov_{\alpha \beta} = [F_{\alpha\beta}^{\rm
lenses}]^{-1},
\end{equation}
from Eq.~(\ref{F_lenses}), and
\begin{equation}
\Cov_{\alpha \beta}^- =
 [F_{\alpha\beta}^{\rm
lenses}]^{-1}\Big|_{\Cov^{XY,\PP}_{\ell\ell'}=0},
\end{equation}
the same construction but with the $XY,\phi\phi$ covariance artificially set to zero.

We can then perform a KL transformation by finding all solutions
to the generalized eigenvalue problem
\be
\label{KL}
	\sum_\beta \Cov_{\alpha \beta} v^{(k)}_\beta 
	= 
	\sum_\beta \lambda^{(k)} \Cov_{\alpha \beta}^{-} v^{(k)}_\beta
	.
\ee
Here $v^{(k)}_\beta$ and $\lambda^{(k)}$ are the KL eigenvectors and eigenvalues. The KL transform of the measurements
\be
\label{kl_modes}
	\ourTh^{(k)} = \sum_\alpha v^{(k)}_\alpha p_\alpha 
\ee
provides a representation that is uncorrelated, or statistically orthogonal, with respect to both covariance matrices since solutions to \eqref{KL} are simultaneously  orthogonal
with respect to the metrics defined by the covariance matrices,
\begin{eqnarray}
\label{orthogonal}
	\Cov_{\ourTh^{(k)}\ourTh^{(l)}}=  \sum_{\alpha\beta} v^{(l)}_\alpha \Cov_{\alpha \beta} v^{(k)}_\beta   &=& \lambda^{(k)} \delta_{kl} , \nonumber\\
	\Cov_{\ourTh^{(k)}\ourTh^{(l)}}^{-} = \sum_{\alpha \beta} v^{(l)}_\alpha \Cov_{\alpha \beta}^{-}
	v^{(k)}_\beta &=& \delta_{kl} .
\end{eqnarray}
We order $\lambda^{(k)}$ to be decreasing with $k$ and hence in the ratio of the variances
between the two, i.e.~the degradation in the constraints due to  $\Cov^{XY,\PP}_{\ell\ell'}$.

The eigenvectors are not necessarily mutually  orthonormal in the ordinary Euclidean sense,
\begin{equation}
\sum_{\alpha} v^{(l)}_\alpha  v^{(k)}_\alpha   \ne \delta_{kl},
\label{euclid}
\end{equation}
as they would be in an ordinary eigenvector or principal component representation (see \S
\ref{sec:PC}).  Consequently, the forward and inverse KL transforms are distinct:
\be
\label{kl_modes}
	p_\alpha = \sum_k w^{(k)}_\alpha \ourTh^{(k)} ,
\ee
where  $w^{(k)}_\alpha$ is the matrix inverse of $v^{(k)}_\alpha$ rather than its transpose.
As a function of the $\alpha$ index, $v^{(k)}_\alpha$ represents {how strongly
individual} $p_\alpha$ contribute to the $k$th KL mode whereas  $w^{(k)}_\alpha$ represents
how the $k$th KL mode is distributed onto the original modes.  They can have very different shapes in $\alpha$.    We always use the forward KL transform and $v^{(k)}_\alpha$ in the following discussion to avoid confusion.

\begin{figure}
\center
\includegraphics[width = 0.49 \textwidth]{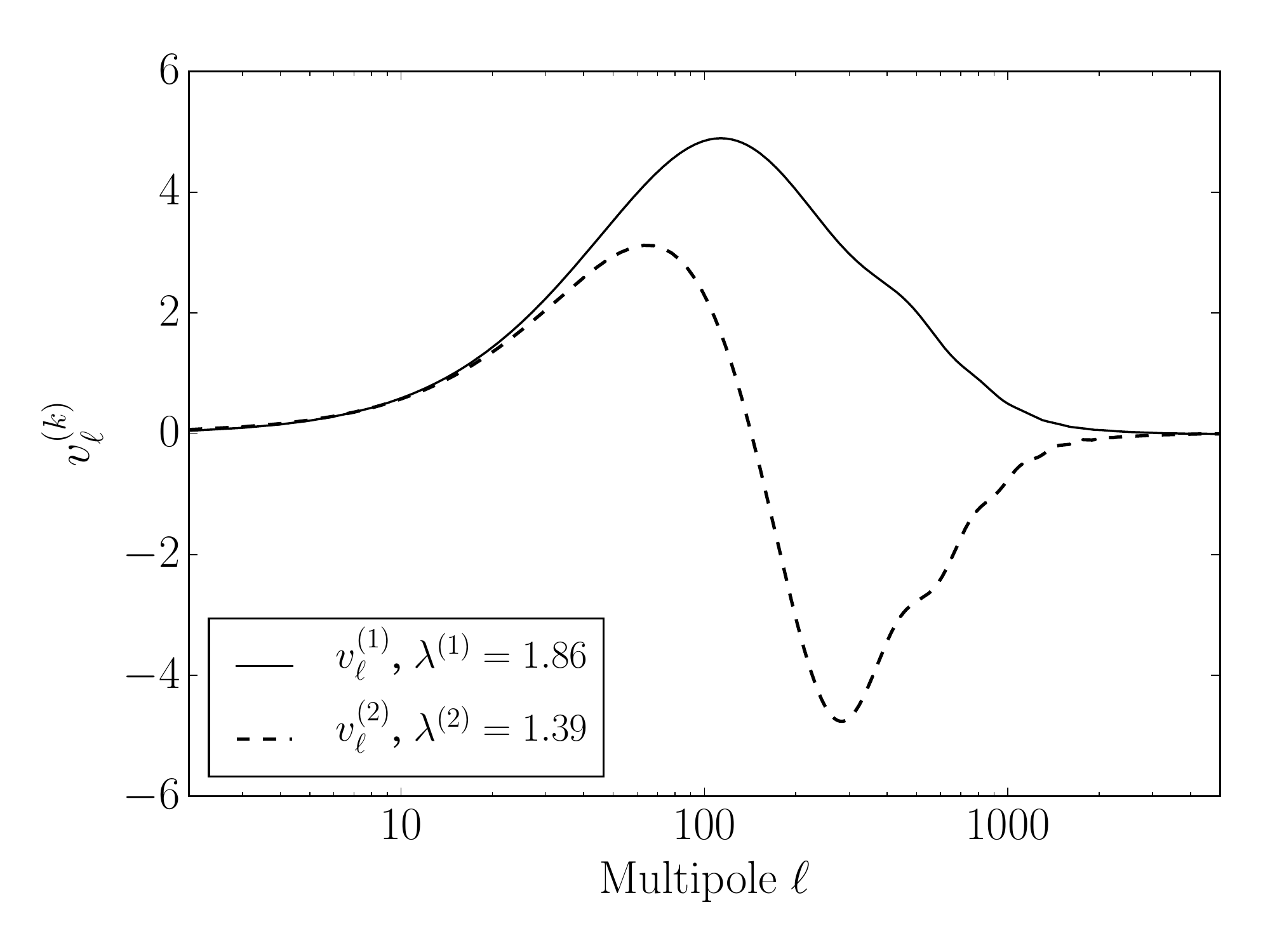}
\caption{KL components of the lensing potential most affected by the covariances
$\Cov^{XY,\PP}$ of CMB fields with the reconstructed lensing potential. 
By neglecting these covariances, constraints on the corresponding amplitude $\Psi^{(k)}$
would be overly optimistic due to double counting of lensing information. 
}
\label{fig:degraded} 
\end{figure}

We find two strongly degraded modes
with
\ba
\lambda^{(1)} &=& 1.86, \\
\lambda^{(2)} &=& 1.39 \nonumber .
\ea
These modes would be 
better constrained if there were no $XY,\PP$
covariances, which agrees with the intuitive expectation that neglecting mutual covariances
would lead to double counting of the lensing information. 
The corresponding eigenvectors $v^{(1,2)}_\alpha$ are plotted in Figure~\ref{fig:degraded}.
All other modes are only mildly affected and have eigenvalues between 0.93 and 1.08. 

We see that measurements of the amplitude of the first mode $\ourTh^{(1)}$ are degraded by almost a factor of two. 
This means that constraint on this mode obtained from the $XY$ lensed power spectra alone
is
 comparable to a constraint  from the reconstructed lensing potential alone but that these two different measurements are highly correlated.
This occurs because both these measurements have their variances
dominated by the sample variance of the lenses. This sample variance is common to both
measurements, which explains why the two variances are comparable and strongly correlated.

Table~\ref{tab:scenarios} summarizes how well we can
constrain $\ourTh^{(1)}$ under various assumptions and provides quantitative justification
of these claims. The first two lines summarize the KL results
-- neglecting $\Cov^{XY,\PP}$ leads to a double counting of
the lensing information and overly tight constraints in the full dataset. 
Instead, we can constrain this mode separately from $\PP$ and $XY$ data with variances
that are both comparable to those of the full dataset.   The $XY$ result is not a trivial consequence of the KL results since the KL modes
are not  specifically constructed to be statistically orthogonal with $XY$ measurements alone.
Because the $XY$ power spectra provide only integrated constraints on $C_\ell^\PP$, we
impose a mild theoretical prior of $\sigma_{p_\alpha} = 1$ to forbid numerical problems and
degeneracies induced by unphysically large features  in $C_\ell^\PP$ (see also \S \ref{sec:PC}).  The  minimum variance unbiased linear estimators of $\ourTh^{(1)}$ from the separate $\phi\phi$ and $XY$ datasets have a correlation coefficient of   0.77, in agreement with values in Table~\ref{tab:scenarios}.

Note that even when considering $XY$ separately, we include all of the internal covariances induced
by lens sample variance.   
Without the non-Gaussian covariances $\mathcal{N}$, $\sigma_{\ourTh^{(1)}}^2$ decreases significantly
and is unphysically smaller than the lens sample variance limit by more than a factor of 3.  Finally we show
that removing all of the non-Gaussian covariances in the full dataset leads to an even more
extreme violation of the lens sample variance limit.

\begin{table}
\caption{Variance of KL  consistency mode $\ourTh^{(1)}$ obtained from various combinations of lensed CMB spectra $XY$ and
lens power spectra $\phi\phi$ measurements and assumptions about their variances and covariance.
}
\label{tab:scenarios}
\begin{tabular}{ccc}
\hline\hline
Dataset & Covariance & $\sigma^2_{\ourTh^{(1)}}$\\
\hline 
$XY, \phi\phi$ &  $\Cov^{XY,\PP}_{\ell\ell'}=0$ & 1.00\\
$XY, \phi\phi$ & Full  & 1.86
\\
\hline
$\phi\phi$ & Full & 1.96\\
$XY$\footnote{\label{note1}With a mild theoretical prior $\sigma_{p_\alpha} = 1$} &
Full & 2.26 \\
\hline
$\phi\phi$ & Sample variance & 1.74\\
$XY$\footref{note1} & Gaussian & 0.52 \\
$XY, \phi\phi$ & Gaussian  & 0.29\\
\hline \hline
\end{tabular}
\end{table}

Because $\ourTh^{(1)}$ is constrained by two independent but strongly correlated
 measurements, these measurements in principle provide an excellent
systematic check on the experimental data that is nearly immune to sample variance
and cosmological parameter uncertainties. This check could be very valuable
in future experiments, which are likely to be foreground and systematics limited: 
comparing $\ourTh^{(1)}$ measured from  power spectra and reconstruction separately could
serve as a simple check on data quality and reconstruction algorithms before performing the delensing operation. 
Identical conclusions are to a lesser degree valid also for $\ourTh^{(2)}$, which could
also serve as a weaker consistency check, but valuable in its own right for reasons we
discuss below.

Next we test the robustness of these results to our assumptions.
The eigenvectors $v^{(k)}$ and corresponding eigenvalues do not change
appreciably if we discard in temperature and polarization information for $\ell < 30$,
discard reconstruction information for $\ell > 3000$, or include polarization information out to $\ell < 5000$.
{Unlike cosmological parameter inferences that involve breaking parameter degeneracies
involving the standard $\Lambda$CDM parameters, $A_s, \tau, \Omega_c h^2$, this consistency
test involves just the lensing information.}  In principle, the development of more 
sophisticated lens reconstruction algorithms beyond the damping tail may in the future allow additional
consistency tests with $XY$ power spectra at $\ell>3000$.  However, this information does not
significantly impact the $\ourTh^{(1)}$ consistency test since it involves lens power on comparably high $\ell$ scales.
The impact of neglecting $C_\ell^{T\phi}, C_\ell^{E\phi}$ should also not be significant, because
unlike $v^{(1,2)}$ they are only significant at the lowest multipoles.

The most important assumption in this construction is that we can independently
consider the information about the
unlensed CMB and the lens power spectra.   While this is a good assumption in the
extended $\Lambda$CDM parameter space for the full data set as demonstrated in Fig.~\ref{fig:model_comparison}, it is less true when considering the
lensed CMB $XY$ spectra alone if spatial curvature is allowed to vary.   Increasing
$\Omega_K$ impacts the  unlensed CMB through $\tilde \Omega_K$ in a manner similar to the
smoothing of the acoustic peaks by lensing \cite{Smith:2006nk}.   Moreover, its impact on lensing through
$p_\alpha(\Omega_K)$  is to decrease the
amplitude of power (see Fig.~\ref{fig:dClpp} below), and so the overall sensitivity to curvature is degraded from what is assumed 
in the independent approximation.  Furthermore, the total impact of curvature on the lensed power spectrum 
becomes nearly degenerate with effects of the neutrino mass \cite{Smith:2006nk}.
On the other hand, $BB$ partially breaks the degeneracy as it is not generated by curvature.

To investigate how severe these degeneracies are in the $XY$ dataset, we compare
forecasted errors on  $\ourTh^{(1,2)}$ with fixed vs.\  marginalized
$\tilde \theta_A$ in 
Table~\ref{tab:unlensed_effect}. As before, we assume a mild theoretical prior
$\sigma_{p_\alpha} = 1$.

When $\tilde{\Omega}_K$ is held fixed, the variances of both  $\ourTh^{(1)}$ and $\ourTh^{(2)}$ 
are negligibly increased by marginalizing the remaining 8 extended $\Lambda$CDM parameters.
When $\tilde\Omega_K$ is also marginalized the variance of  $\ourTh^{(1)}$ changes only by 
$\sim 10\%$ but that of  $\ourTh^{(2)}$ is close to doubled.
This mirrors the fact that changing $\ourTh^{(1)}$ changes $BB$ significantly more
--
relative to the rest of the observables --  than $\ourTh^{(2)}$ does and cannot be mimicked by
curvature in the unlensed spectra.
We conclude that  $\ourTh^{(1)}$ provides a robust consistency test for lensing in the full
$\Lambda$CDM+$w$+$\Omega_K$+$\sum m_\nu$ context whereas
inconsistencies in
 $\ourTh^{(2)}$ between $XY$ and $\phi\phi$ measurements may indicate a finite spatial curvature.
Violations of consistency in $\ourTh^{(1)}$ would indicate systematics and foregrounds in the measurement or
 new physics at recombination that mimics the effect of lensing.  Either of these possibilities would lead to 
 incorrect cosmological inferences and 
 complicate delensing of the CMB if not discovered beforehand.

This relationship between  lensing  and curvature effects in the unlensed
spectrum also leads to the small
difference between the full Fisher forecast and the independent lensing information model 
in Figure~\ref{fig:model_comparison}
which we discuss further in Appendix \ref{sec:appendix}.

\begin{table}
\caption{Variance of KL consistency modes $\Psi^{(1,2)}$ obtained from $XY$ lensed CMB power
spectra alone with and without unlensed CMB parameters $\tilde
\theta_A$  marginalized.\footnote{
With mild theoretical prior $\sigma_{p_\alpha} = 1$.}} 
\label{tab:unlensed_effect}
\begin{tabular}{c|cc}
\hline\hline
& $\sigma^2_{\ourTh^{(1)}}$ & $\sigma^2_{\ourTh^{(2)}}$ \\[0.5ex]
\hline
all $\tilde \theta_A$ fixed & 2.26 & 4.13\\
8  marginalized, $\tilde \Omega_K$ fixed   & 2.27 & 4.35\\
all  $\tilde \theta_A$ marginalized  & 2.52 & 7.34\\
\hline\hline
\end{tabular}
\end{table}

\subsection{Principal component implementation}
\label{sec:PC}

The consistency check discussed in  \S \ref{sec:consistency} involves measuring the KL consistency
parameter  $\Psi^{(1)}$
from the CMB $XY$ power spectra alone.    There are practical obstacles to implementing this measurement
given the many ill-constrained modes that compose the full lensing power spectrum $C_\ell^{\phi\phi}$ through
$p_\alpha$.   Furthermore, with just $XY$ measurements alone, curvature $\Omega_K$ mildly violates
the assumption that the unlensed CMB parameters can be independently
 extracted from the lensed CMB as
discussed in the previous section.  
A full assessment will require  going
beyond the Fisher approximation with 
validation on numerical simulations which we postpone to  a future work.   In this section, we take
the first steps toward this goal by re-examining the lensing principal component decomposition 
introduced in Ref.~\cite{Smith:2006nk}.   A small set of these parameters completely characterizes the lensing information in the $XY$ data and can be measured jointly 
with those controlling the unlensed parameters $\tilde \theta_A$, with or without curvature.

The forecasted covariance matrix of the $p_\alpha$ 
lensing parameters  measured by  $XY$ power spectra is given by  the inverse Fisher
matrix \eqref{F_lenses}, omitting $\PP$ in the sum. The orthonormal eigenvectors
$K^{(i)}_\alpha$ of this matrix represent an alternate basis for the measurements
\be
	\Theta^{(i)} = \sum K^{(i)}_\alpha p_\alpha,
\ee
that yield uncorrelated parameters, rank ordered by their variance, in principle.   By keeping only the  eigenvectors that are predicted to have low variance, we  can measure the relevant information with a much smaller set of principal components (PCs).  
Note that this differs from the KL basis in that it rank orders modes
by total variance from $XY$ rather than by whether the joint measurements are noise or lens sample variance dominated.

The efficiency of the PC approach depends on the number of components needed
to completely characterize the relevant information.
 In
our case, we find eigenvalues
\ba
	10^3 \lambda &=& 1.0, 4.0, 12, 19, 93, \dots ,
\ea
which indeed shows that relative importance of the components decreases rapidly and hits
the $\sigma_{p_\alpha} = 1$ prior shortly thereafter.

\begin{figure}
\center
\includegraphics[width = 0.49 \textwidth]{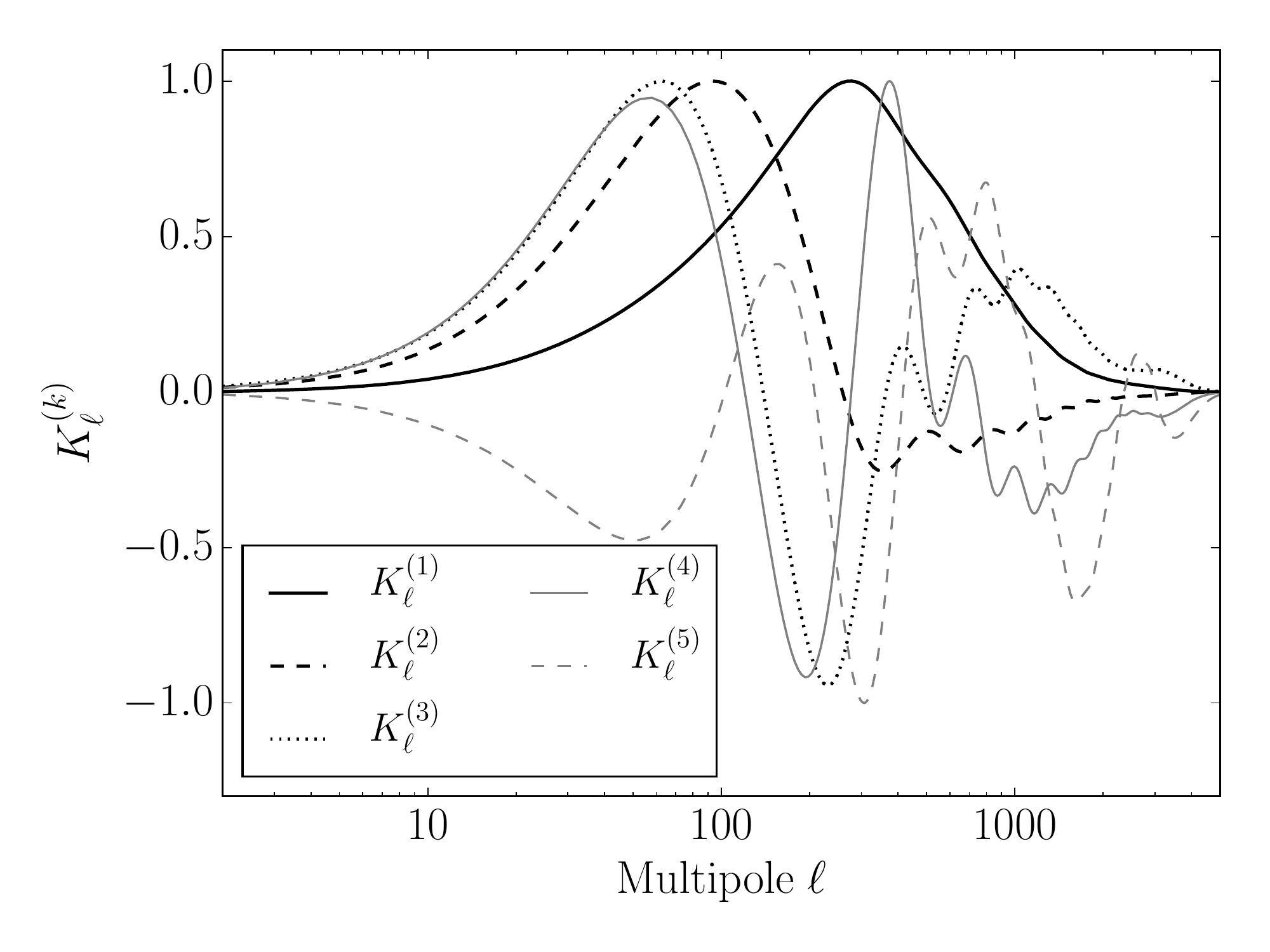}
\caption{Five principal components $K^{(i)}_\ell$ of the lensing potential
best measured by the lensed power spectra.}
\label{fig:pca}
\end{figure}

The five most important components are shown in Figure~\ref{fig:pca}.
The
low order modes peak where the lenses have their largest impact on $XY$ and the higher
modes are increasingly oscillatory, because they have to be orthogonal to the more
important eigenmodes.

It is sufficient to keep only several principal components to characterize the impact of
cosmological parameters or the KL consistency modes completely. 
Specifically, the mode $\ourTh^{(1)}$ can be faithfully constructed from $XY$ 
measurements of the 5 lowest order  PC components with the dominant contributions from the
first two. We have explicitly checked
that truncating the remaining components has no significant effect on the error analysis, for example
as displayed in  Table~\ref{tab:unlensed_effect}.
Because of the truncation, the $\sigma_{p_\alpha} = 1$ prior plays little role and may be omitted.
This construction therefore provides a practical means of measuring  $\ourTh^{(1)}$ in the presence
of the many unconstrained but unphysical modes.

We can also measure these $\Theta^{(i)}$ modes with lensing reconstruction and check consistency
between $XY$ and $\phi\phi$  directly in PC space.   The results are summarized
in Table~\ref{tab:pca}.  Although the first mode is equally well constrained by $XY$ and
$\phi\phi$ measurements,
it does not produce as sharp a consistency test as  $\ourTh^{(1)}$. The reason is that lens
sample variance only contributes less than $\sim 2/3$ of the variance of either
measurement and their results can therefore
differ due to the remaining noise variance.   Higher modes are even less sample variance limited in $XY$.
This mainly reflects the higher $\ell$ weight in the PC components compared with  $\ourTh^{(1)}$.
We can interpret $\ourTh^{(1)}$ as essentially the linear combination of $\Theta^{(1)}$ and
$\Theta^{(2)}$ that best isolates the low $\ell$, lens sample variance limited information.

Finally, while in this work we mainly focused on lensing information which is redundant, these results
imply that the lensed 
$XY$ CMB 
power spectra actually  improve constraints on lensing potential above roughly $\ell \sim 500$ 
(see Tab.~\ref{tab:pca}). 
   In cosmological parameter errors this improvement is hidden
because of the degeneracies with \LCDM parameters as we discuss next.

\begin{table*}
\caption{Variance of $\Theta^{(i)}$ obtained from various datasets under various
assumptions about covariances of the data and noise. }
\label{tab:pca}
\begin{tabular}{ccccccc}
\hline\hline
Dataset & Covariance 
& $10^3 \sigma_{\Theta^{(1)}}^2$
& $10^3 \sigma_{\Theta^{(2)}}^2$
& $10^3 \sigma_{\Theta^{(3)}}^2$
& $10^3 \sigma_{\Theta^{(4)}}^2$
& $10^3 \sigma_{\Theta^{(5)}}^2$
\\[0.5ex]
\hline 
$XY$\footnote{\label{note1}With a mild theoretical prior $\sigma_{p_\alpha} = 1$} & Full
& 1.0 & 4.0 & 12 & 18 & 85\\
$\phi\phi$ & Full
& 1.0 & 2.4 & 5.7 & 8.2 & 30\\
\hline
$\phi\phi$ & Sample variance
& 0.66 & 2.0 & 1.2 & 0.8 & 0.4\\
\hline
 \hline
\end{tabular}
\end{table*}

\subsection{Parameter constraints revisited}

The KL analysis exposes the fact that there is one mode which is nearly equally well
measured by CMB power spectra $XY$ and lensing reconstruction $\phi\phi$ that reflects a large
portion of the nearly lens sample dominated information on $C_\ell^{\phi\phi}$ at low 
$\ell$.   Our PC analysis highlights the fact that the decrease in lens sample variance
at higher $\ell$ means that despite being the highest in signal to noise, this consistency
mode carries only a portion of the total information from lensing on the overall 
amplitude of the lensing spectrum. Furthermore, as shown in
Fig.~\ref{fig:param_constraints}, the constraints from the overall amplitude of the lensing
power spectrum on the $\Lambda$CDM extensions is limited by degeneracies since the  $\Lambda$CDM parameters $A_s$ (implicitly $\tau$) and
$\Omega_c h^2$ also affect $p_\alpha$, further reducing the impact of lens sample covariance.
It is when the low $\ell$ {lensing} information strongly breaks a parameter degeneracy
that the impact of the $\Cov^{XY,\PP}$ covariance is seen.

\begin{figure*}
\center
\includegraphics[width = 0.49 \textwidth]{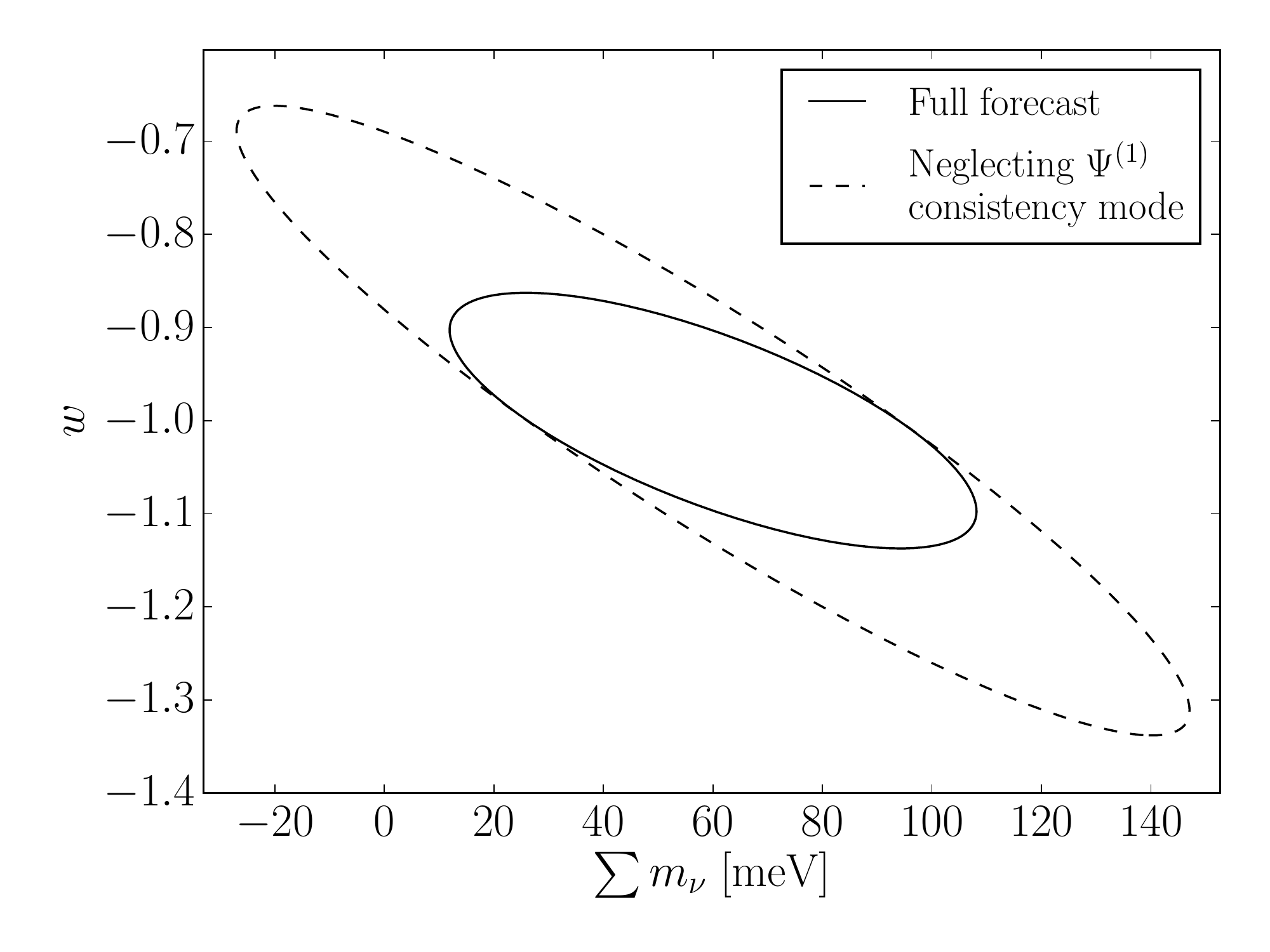}
\includegraphics[width = 0.49 \textwidth]{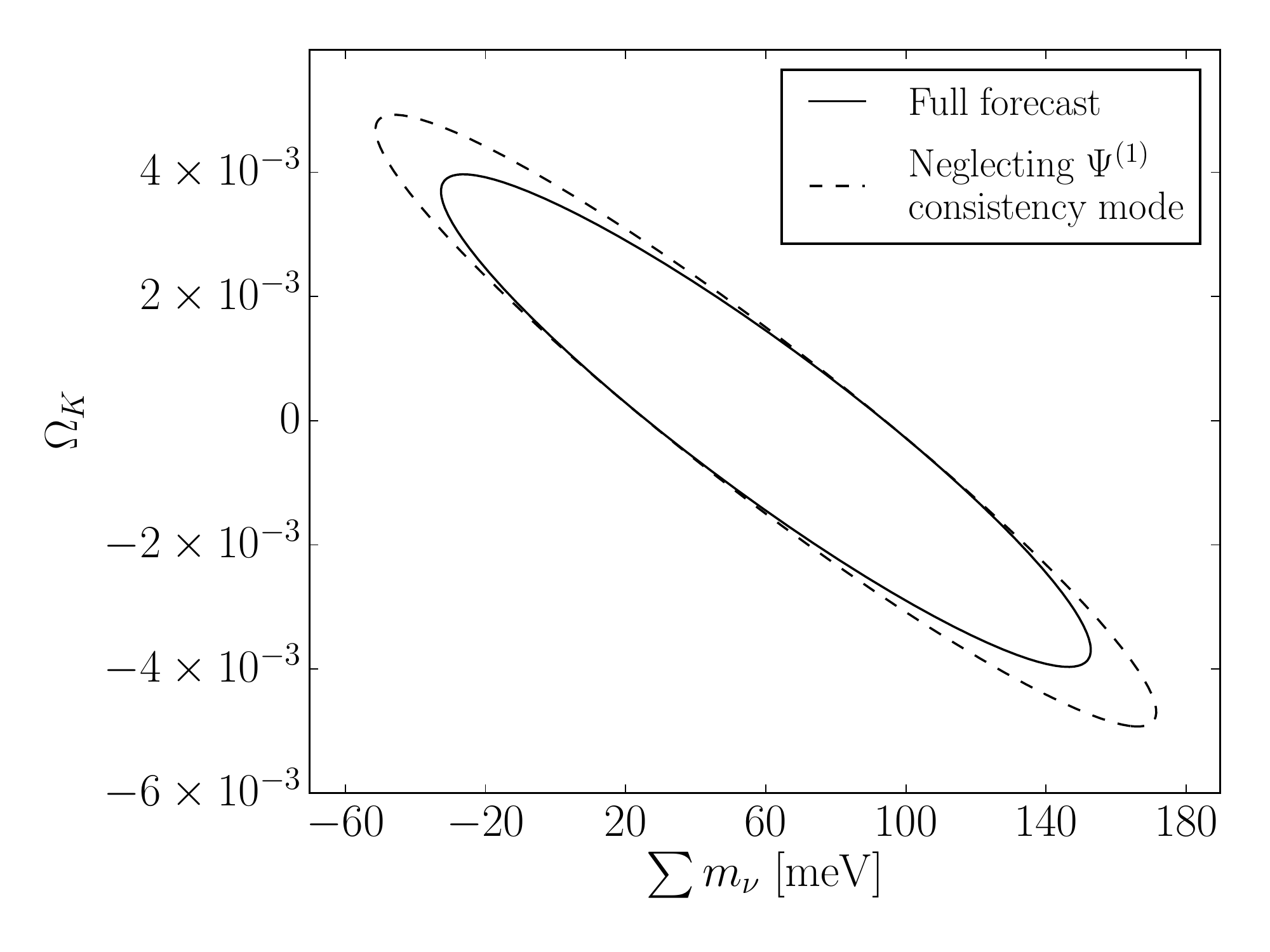}
\caption{
Impact of eliminating the lens information associated with the KL consistency
mode $\Psi^{(1)}$ (dashed line).  Solid lines represent the full Fisher
 forecast  from Fig.~\ref{fig:param_constraints}.   The 
 consistency mode carries a substantial amount of the total information especially in 
 cases where low $\ell$ lens information breaks  parameter degeneracies. 
}
\label{fig:psi_1_effect}
\end{figure*}

In Figure~\ref{fig:psi_1_effect} we show how the parameter
constraints would change if we neglect the information carried by the $\Psi^{(1)}$ consistency mode. In both the $w-\sum m_\nu$ and $\Omega_K -\sum m_\nu$ cases, the impact is mainly
in the degenerate direction but is only dramatic in the former.  
The impact of  $\Cov^{XY,\PP}$ shown in Fig.~\ref{fig:param_constraints} can be
understood from this result since the information on $\Psi^{(1)}$ is essentially double 
counted if this covariance is neglected.

\begin{figure}
\center
\includegraphics[width = 0.49 \textwidth]{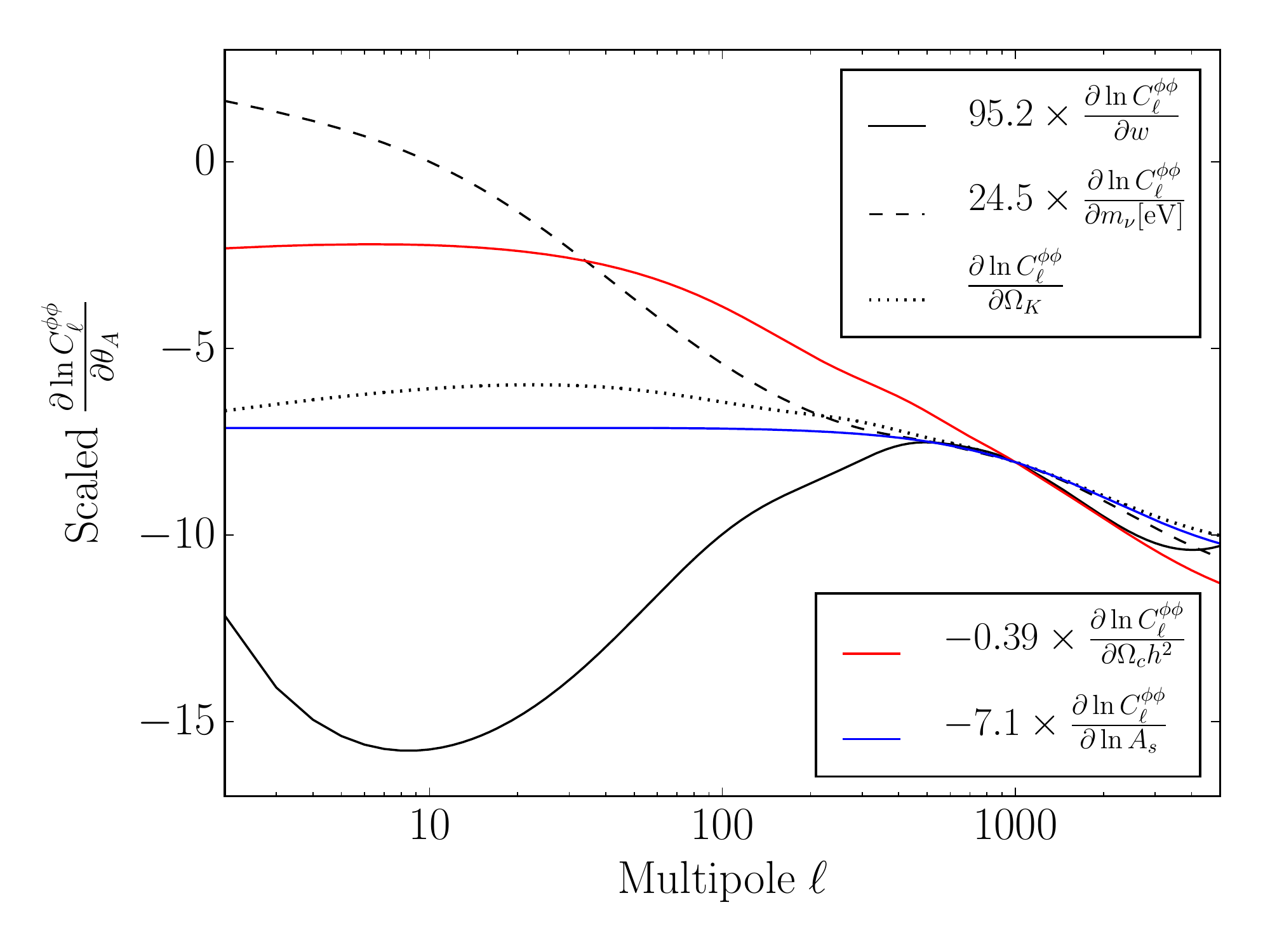}
\caption{Derivatives of $\ln C_\ell^{\phi\phi}$ with respect to cosmological parameters
$w, \sum m_\nu, \Omega_K, \Omega_c h^2, \ln A_s$ normalized at $\ell=1000$ to highlight degeneracies.
These derivatives are taken at fixed acoustic scale $\theta$.
}
\label{fig:dClpp}
\end{figure}

We can further understand the different parameter behaviors by examining the impact of
parameters on $p_\alpha$ or $\ln C_\ell^\PP$ (see Fig.~\ref{fig:dClpp}).
Although the measurements determine the amplitude of the $C_\ell^\PP$ well at
$\ell \gtrsim 500$, they are unable to separate out the contributions from the various 
cosmological parameters.    In particular, linear combinations of $\ln A_s$ and $\Omega_c h^2$
can mimic the impact of the extended $\Lambda$CDM parameters \cite{Smith:2006nk}.
Therefore, while the best constrained direction in the 2-dimensional extended spaces correspond to combinations
of the parameters that coherently change  $C_\ell^\PP$ at $\ell \gtrsim 500$, the constraint itself
is limited by how well $\ln A_s$ and $\Omega_c h^2$ are measured not by how well
$C_\ell^\PP$ is measured (see Fig.~\ref{fig:param_constraints}).  The degenerate or 
worst constrained direction corresponds to when the parameter variations cancel in their
effect.  

At $\ell \lesssim 500$ the degeneracy between $w$ and
$\Omega_K$ or $\sum m_\nu$ observed at high $\ell$ starts to break, which allows us to
meaningfully constrain also the perpendicular direction in the parameter space. For
$\Omega_K$ and $\sum m_\nu$ this degeneracy breaking is noticeably weaker, especially at
$\ell \gtrsim 50$.  Given the large sample variance associated with the lowest multipoles, the limiting
source of information in the degenerate direction in the $\Omega_K-\sum m_\nu$ plane comes from the
unlensed CMB rather than the lensing information.   Hence, 
the effect of lens sample covariance is smaller in this case.

Finally for these issues that relate to parameter degeneracies, it is important to remember that
external information from measurements 
beyond the CMB, for example from baryon acoustic oscillations, can break these
degeneracies and allow more of the information on $C_\ell^{\phi\phi}$ that our analysis uncovers
to be used for parameter constraints.

\section{Discussion}
\label{sec:discuss}

The lensing observables from the two and higher point statistics of the temperature and
polarization fields are intrinsically correlated because they are lensed by the same
realization of structure  between last scattering and the observer.    While currently these
observables are noise variance limited, in the future they are expected to be lens sample
variance limited.   
When jointly analyzing these observables, it will then be important to take these correlations into account
 both to prevent double counting of information and because they provide important 
consistency checks that are immune to sample variance, the chance fluctuations in the lenses.  

In this work, we study a simple analytical model that consistently incorporates the lens sample
covariance between CMB power spectra and lens reconstruction from higher point information.
This covariance model can be  employed for cosmological parameter estimation to
build the lens sample variance piece of the 
likelihood function as well as Fisher forecasts  for future experiments.
 
While there is only a small effect on parameter errors of the covariances between the reconstructed lensing potential
and the lensed power spectra in the \LCDM and even the extended \LCDM context,
parameter errors, degeneracies and non-lensing information mask the full impact of the
covariance.  

To  better expose this impact, we work in an approximation where information in the
unlensed CMB power spectrum and the lensing potential
$C_\ell^\PP$ are considered independently. 
Using a  Karhunen-Lo{\`e}ve  analysis, we identify one mode in 
$C_\ell^\PP$ that in the future should be nearly lens sample variance limited 
using either lensed power spectra or lensing reconstruction 
and hence nearly perfectly covaries between the two.   If this covariance
is not taken into account then information on this mode will be double counted.
This mode peaks at somewhat lower multipole than the bulk of the information on the lensing power
spectrum due to the larger signal versus noise variance there.  

This mode can be measured separately through  lens reconstruction and lensed CMB power spectra
with the help of a principal component decomposition of the latter.
  Notably, inconsistency between the measurements cannot
be explained by chance lens realizations or parameter variations, 
and is immune to ambiguities
due to $\tau$, the optical depth to reionization. Instead, violations  could indicate systematics,
lens reconstruction errors, foregrounds or
new physics at recombination which changes the unlensed power
spectra, including the $BB$ power spectrum,  in ways degenerate with lensing.
They would then lead to incorrect cosmological
inferences and delensing if not taken into account.

The identification of this mode also explains the impact of covariances between the reconstructed lensing potential
and the lensed power spectra on parameter constraints.    There is only a small effect within
the \LCDM model as these parameters are well constrained even without lensing.  
The impact of covariance is mainly seen when measurements of the low $\ell$ lensing power spectrum
are useful in breaking parameter degeneracies in interpreting the measurements at higher $\ell$.
Specifically, for $w$ and $\sum m_\nu$,  the consistency mode has a strong impact on parameters
and hence its double counting would lead to constraints overly optimistic by $\sim$ 20\%.

There is a second combination of $C_\ell^\PP$ with similar properties, however there the
correlation is weaker.  Despite being weaker, statistically significant violations of consistency  
in this
mode are interesting since they may indicate nonzero spatial curvature as it has
 similar effects on the unlensed CMB as lensing.  

\medskip

While this work was in preparation, a similar analytic approach to modelling covariances was
compared against numerical simulations 
\cite{Peloton:2016kbw}. That model was found to work
 well after realization-dependent noise subtraction. As can be seen from their Figs.\ 3 and 4, 
these subtractions affect mostly correlations with lensing power spectra above $\ell \sim
1000$ and would be hidden by reconstruction noise in our approach.   They also show that
the other trispectrum
 terms to the covariance, which we neglect, are  subdominant. Potentially more
troublesome is their finding that there are some differences between the analytical model and
simulations, especially in $\Cov^{BB,\PP}$ at low $\ell_{BB}$ which they claim
{appears} to be statistically
significant \cite{Peloton:private}.   If confirmed, then our analysis implicitly assumes that such additional
effects can be modeled without breaking our consistency relations -- in essence that both lensed CMB and 
reconstruction can measure this {consistency} mode to nearly the lens sample variance
limit.    More generally, this consistency mode can be used to search for unaccounted for
systematics in lens reconstruction. We intend to study
these issues and quantify their impact in a future work.

\acknowledgements{
We thank Chen He Heinrich, Alessandro Manzotti and Julien Peloton for useful discussions.
This work was
supported by U.S.~Dept.\ of Energy contract DE-FG02-13ER41958 and in part by the Kavli
Institute for Cosmological Physics at the University of Chicago through grant NSF
PHY-1125897 and an endowment from the Kavli Foundation and its founder Fred Kavli.  WH was
additionally supported by the Kavli Institute for Cosmological Physics at the University
of Chicago through grants NSF PHY-0114422 and NSF PHY-0551142 and NASA ATP NNX15AK22G and thank the Aspen Center for Physics, which is supported by National Science Foundation grant PHY-1066293, 
where part of this work was completed.
ABL thanks CNES for financial support through its postdoctoral programme and KICP, where this work was initiated, for its visitor program.
We acknowledge use of the CAMB software package. This work was completed in part with
resources provided by the University of Chicago Research Computing Center.
}

\appendix

\section{Simple forecast methods}
\label{sec:appendix}

In this appendix we compare various Fisher matrix approaches of how to estimate parameter constraints,
including the standard calculation which uses the full analytical
covariance matrix \eqref{full_covariance}.
  We also introduce a new
forecasting approach, which we call the Simple Lensing Approximation (SLA), that is very accurate in
predicting parameter constraints from CMB data only and does
not require calculation of the full covariance matrix. 

\begin{figure*}
\center
\includegraphics[width = 0.49 \textwidth]{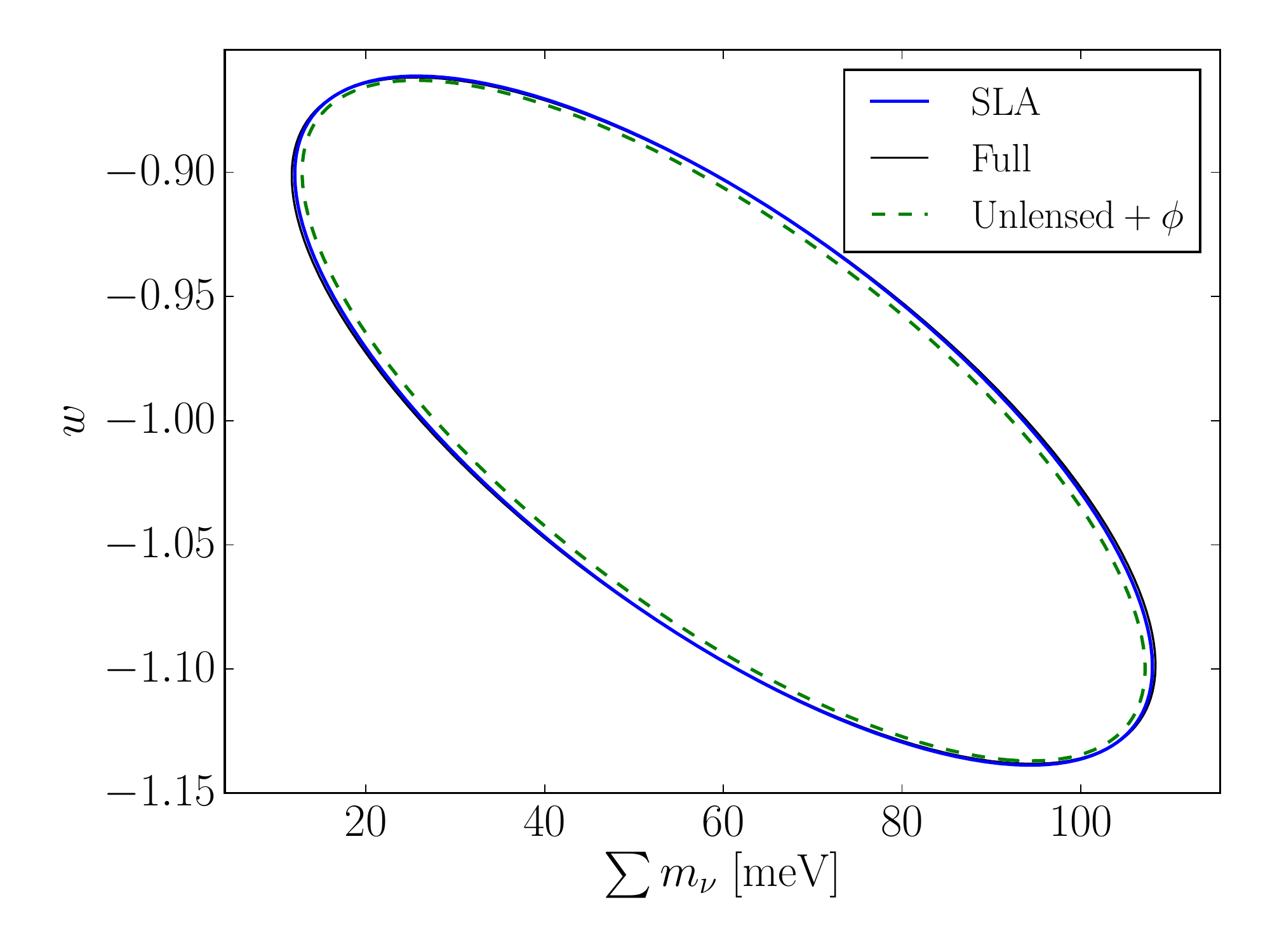}
\includegraphics[width = 0.49 \textwidth]{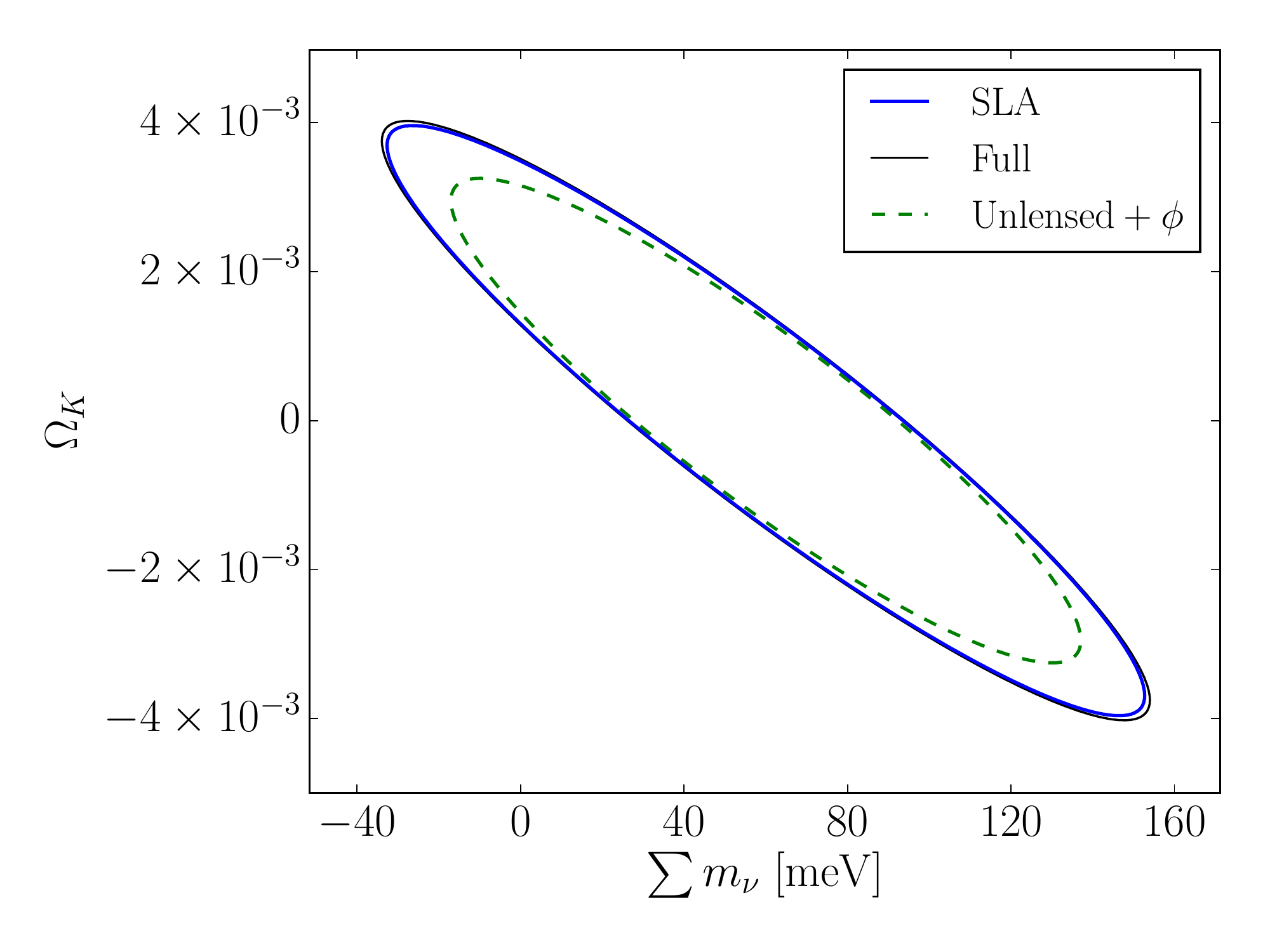}
\caption{
New SLA forecasting approximation (solid blue) compared with the full forecast (black solid) 
and the  frequently used
approach of using the unlensed spectra and $\phi$ lensing power spectrum (green dashed).   The unlensed approach
is simple but errs in assuming the unlensed spectra are directly observable.  The SLA approach employs the
lensed CMB spectra but omits both their lensing information and  covariances making it both accurate and simple.
}
\label{fig:more_model_comparison}
\end{figure*}

A frequently used approach to avoid double counting of the lensing information is
to derive parameter constraints from the unlensed  $\tilde X \tilde Y$
CMB power spectra
and the reconstruction of the lensing potential assuming Gaussian statistics in each.
These constraints are equivalent to assuming that complete delensing in the CMB maps is possible,
that it does not alter their noise properties and that no extra information on the lensing
beyond reconstruction can be recovered from the $XY$ power spectra. 
In the main text we have
seen that while the lensing information in $XY$ is substantial, it is largely redundant with
reconstruction or limited by parameter degeneracies.    For this reason, this approximation works fairly well in the $w-\sum m_\nu$ plane.  
However, as seen from
Figure~\ref{fig:more_model_comparison}, this approximation noticeably underestimates the errors on curvature since its effect on the unlensed spectrum and lensing work in 
opposite directions in the smoothing of the peaks,  degrading the overall  curvature sensitivity in the
lensed CMB power spectra.

This problem is largely fixed by our  independent lensing information model of
Eq.~\eqref{simple_model} which is shown in Fig.~\ref{fig:model_comparison}.    In this model, the information from the unlensed
power spectra is still considered as separate from that of the lens spectrum but 
the observable is the lensed $XY$ spectrum and lens sample covariance is taken into
account in the covariances of observables.   The drawback is that to make forecasts,
the cumbersome
lens sample covariance matrix must be carried through all pieces of the construction.

We can combine the virtues of these two approaches in a new simple forecasting method, dubbed SLA,
if all that is desired {are} parameter forecasts in the extended \LCDM space from the CMB alone.
Namely, we can avoid double counting of the lens information by dropping the
lens information in the $XY$ power spectra and along with it the non-Gaussian covariances
induced by lensing.  Importantly, we still use the lensed $XY$ power spectra and not the unlensed $\tilde X\tilde Y$ power spectra as the observables.
Specifically,
\be
\label{simple_model2}
	F^\mathrm{SLA}_{AB} = F_{AB}^{\mathrm{unl,SLA}} + \sum_{\alpha,\beta} 	
	\frac{\partial p_\alpha}{\partial \theta_A}
 F_{\alpha\beta}^{\mathrm{lenses,SLA}}\frac{\partial p_\beta}{\partial \theta_B}
\ee
where we continue to assume Gaussian lens reconstruction noise as in the main text,
\ba
\label{F_lenses2}
	{F_{\alpha\beta}^{\mathrm{lenses,SLA}} =} 
	&&\sum_{ \ell }
	\frac{\partial C^{\PP}_\ell}{\partial p_\alpha} 
	\(\mathcal{G}^{\PP,\PP}_{\ell \ell}\)^{-1}
	\frac{\partial C^{\PP}_{\ell}}{\partial p_\beta}.
\ea
and omit any lensing information in the $XY$ power spectra.
The conceptual difference from Eq.~(\ref{simple_model}) is that when evaluating
the unlensed Fisher matrix we assume Gaussian statistics,
\ba
\label{F_unl2}
	F_{AB}^{\mathrm{unl,SLA}} &=& \sum_{	\substack{
		\text{$ \ell$}
		\\
		\text{$XY, WZ$}
	}} 
	\frac{\partial C^{XY}_\ell}{\partial \tilde \theta_A} 
	\(\mathcal{G}^{XY,WZ}_{\ell \ell}\)^{-1}
	\frac{\partial C^{WZ}_{\ell}}{\partial \tilde \theta_B} \nonumber. \\
\ea
As before, derivatives in \eqref{F_lenses2} should be evaluated at fixed unlensed power
spectra while derivatives in \eqref{F_unl2} should be evaluated at fixed lensing
potential.

We show in Fig.~\ref{fig:more_model_comparison} that this approximation provides
simple but highly accurate constraints even when curvature is involved.   
In fact it performs slightly better
than the independent approximation of the main text in that it allows lensing to recover
information that would otherwise be lost to the non-Gaussian correlations between multipole
moments in the $XY$ power spectra.
 On the
other hand, this simple forecast scheme ignores the fact that the $XY$ power spectra
provide strong constraints on the lensing power spectra at low multipole  that serve as
consistency checks against reconstruction measurements and provide additional constraints
at high lens multipole when parameter degeneracies are broken by external measurements.
This is especially true beyond the $\ell<3000$ limit for polarization measurements tested here but there the astrophysical uncertainties in modeling lenses in the nonlinear regime also limit cosmological parameter information.

\bibliography{lenscov}

\end{document}